\documentclass[11pt]{article} %

\usepackage[margin=1.1in]{geometry} %
\usepackage{mdpaper}%
\usepackage[separate]{mdtheorems}%
\usepackage{authblk}

\gdef\gitISODate{2018-01-10}
\gdef\gitTag{3.1.2}
\makeatletter
\newcommand\gitDate{%
    \expandafter\@gitDate\gitISODate\relax
}
\def\@gitDate#1-#2-#3\relax{%
    \begingroup
     \year=#1\relax
     \month=#2\relax
     \day=#3\relax
     \today
    \endgroup
}
\makeatother

\title{\normalsize\bfseries\uppercase{Minimum Distance Approach to Inference with Many
    Instruments}\thanks{Previously circulated under the title ``Integrated
    Likelihood Approach to Inference with Many Instruments.'' I am deeply
    grateful to Guido Imbens and Gary Chamberlain for their guidance and
    encouragement. I also thank Joshua Angrist, Adam Guren, Whitney Newey, Jim
    Stock, Peter Phillips and participants at various seminars and conferences
    for their helpful comments and suggestions.}}%

\author{Michal Kolesár\thanks{Correspondence to: Department of Economics, Julis
    Romo Rabinowitz Building, Princeton University, Princeton, NJ 08544. Electronic correspondence:
    \nolinkurl{mkolesar@princeton.edu}. }}%
\affil{\vspace{-1em}\small{Department of Economics and Woodrow Wilson School, Princeton University}}

\date{\normalsize{This version \gitTag, {\gitDate}\\First circulated: November 2012.}}%

\newcommand{\h}[2]{\hat{#1}_{\text{\an{#2}}}} %

\newcommand{\re}{\text{\an{re}}}%
\newcommand{\li}{\text{\an{li}}}%
\newcommand{\md}{\text{\an{md}}}%
\newcommand{\emd}{\text{\an{emd}}}
\newcommand{\umd}{\text{\an{umd}}}
\newcommand{\liml}{\text{\an{liml}}}%
\newcommand{\Zp}{Z}%
\newcommand{\Ni}{k_{n}}%
\newcommand{\Ne}{\ell_{n}}%
\newcommand{\nus}{n-\Ni-\Ne}%
\newcommand{\alphae}{\alpha_{\ell}}%
\newcommand{\alphai}{\alpha_{k}}%
\newcommand{\alphanu}{1-\alphai-\alphae}%
\newcommand{\pifs}{\pi_{2,n}} 
\newcommand{\piss}{\pi_{1,n}} 
\newcommand{\pisse}{\psi_{1,n}} 
\newcommand{\pifse}{\psi_{2,n}} 
\newcommand{\lambdan}{\lambda_{n}} %
\newcommand{\omegan}{\omega_{n}} %
\newcommand{\etan}{\eta_{n}} %
\newcommand{\Qs}{Q_{\mathcal{S}}} 
\newcommand{\hQs}{\hat{Q}_{\mathcal{S}}} 
\newcommand{\Qt}{Q_{\mathcal{T}}} 

\newcommand{\ns}{\text{ns}}

\newcommand{\Psin}{\Psi_{n}} 
\newcommand{\Pin}{\Pi_{n}} 

\newcommand{\linv}{\mathcal{L}_{\text{\an{inv}},n}}

\newcommand{\Qf}{Q_{n}}
\newcommand{\Pm}{P_{n}}
\newcommand{\Mm}{M_{n}}
\newcommand{\Em}{U_{n}}
\newcommand{\ei}[1]{u_{#1 n}}

\begin{document}
\maketitle
\begin{abstract}\setstretch{1.2}
  I analyze a linear instrumental variables model with a single endogenous regressor and many instruments. I use invariance arguments to construct a new minimum distance objective function. With respect to a particular weight matrix, the minimum distance estimator is equivalent to the random effects estimator of \citet{ci04}, and the estimator of the coefficient on the endogenous regressor coincides with the limited information maximum likelihood estimator. This weight matrix is inefficient unless the errors are normal, and I construct a new, more efficient estimator based on the optimal weight matrix. Finally, I show that when the minimum distance objective function does not impose a proportionality restriction on the reduced-form coefficients, the resulting estimator corresponds to a version of the bias-corrected two-stage least squares estimator. I use the objective function to construct confidence intervals that remain valid when the proportionality restriction is violated.

  \vspace{1ex}\noindent\textbf{Keywords:}\ Instrumental Variables, Minimum Distance, Incidental
  Parameters, Random Effects, Many Instruments, Misspecification, Limited
  Information Maximum
  Likelihood, Bias-Corrected Two-Stage Least Squares.
\end{abstract}

\clearpage

\section{Introduction}
This paper provides a principled and unified way of doing inference in a linear
instrumental variables model with a single endogenous regressor and
homoscedastic errors in which the number of instruments, $\Ni$, is potentially
large. To capture this feature in asymptotic approximations, I employ the many
instrument asymptotics of \citet{kunitomo80}, \citet{morimune83}, and
\citet{bekker94} that allow $\Ni$ to increase in proportion with the sample
size, $n$. I focus on the case in which collectively the instruments have
substantial predictive power, so that the concentration parameter grows at the
same rate as the sample size. I make no assumptions about the strength of
individual instruments. I allow the rate of growth of $\Ni$ to be zero, in which
case the asymptotics reduce to the standard few strong instrument asymptotics.

The presence of many instruments creates an incidental parameters problem
\citep{ns48}, as the number of first-stage coefficients, $\Ni$, increases with
the sample size. To directly address this problem, I use sufficiency and
invariance arguments together with an assumption that the reduced-form errors
are normally distributed to reduce the data to a pair of two-by-two matrices. In
the absence of exogenous regressors, the first matrix can be written as
$T= \bigl(y\;\;x\bigr)'P_{Z}\bigl(y\;\;x\bigr)/n$, where $P_{Z}$ is the
projection matrix of the instruments $Z$, and $y$ and $x$ are vectors
corresponding to the outcome and the endogenous regressor. The second matrix,
$S=\bigl(y\;\;x\bigr)'(I_{n}-P_{Z})\bigl(y\;\;x\bigr)/(n-\Ni)$, where $I_{n}$ is
the identity, corresponds to an estimator of the reduced-form covariance matrix.
This solves the incidental parameters problem because the distribution of $T$
and $S$ depends on a fixed number of parameters even as $\Ni\to\infty$: it
depends on the first-stage coefficients only through the parameter $\lambdan$, a
measure of their collective strength.

I then drop the normality assumption and use a restriction on the first moment
of $T$ implied by the model to construct a minimum distance (\md) objective
function. This restriction follows from the property of the instrumental
variables model that the coefficients on the instruments in the first-stage
regression are proportional to the coefficients in the reduced-form outcome
regression. I use this \md\ objective function to derive three main results.

First, I show that minimizing the \md\ objective function with respect to the
optimal weight matrix yields a new estimator of $\beta$, the coefficient on the
endogenous regressor, that exhausts the information in $T$ and $S$. In
particular, this efficient \md\ estimator is asymptotically more efficient than
the limited information maximum likelihood (\liml) estimator when the
reduced-form errors are not normal. Standard errors can easily be constructed
using the usual sandwich formula for asymptotic variance of minimum distance
estimators.\footnote{Software implementing estimators and standard errors based
  on the \md\ objective function is available at
  \url{https://github.com/kolesarm/ManyIV}.} The \md\ approach thus gives a
simple practical solution to the many-instrument incidental parameters problem.

Second, I compare the \md\ approach to that based on the invariant
likelihood---the likelihood, under normality, based on $T$ and $S$. I show that,
when combined with a particular prior on $\lambdan$, the likelihood is
equivalent to the random-effects (\re) quasi-maximum likelihood of \citet{ci04},
and that maximizing it yields \liml. Therefore, the random-effects estimator of
$\beta$ is in fact equivalent to \liml. Furthermore, I show that the \re\
estimator of the model parameters also minimizes the \md\ objective function
with respect to a particular weight matrix. This weight matrix is efficient
under normality, but not in general.

Third, I consider minimum distance estimation that leaves the first moment of
$T$ unrestricted. This situation arises, for instance, when the instrumental
variables model is used to estimate potentially heterogeneous causal effects, as
in \citet{ai95}. When the causal effect is heterogeneous, the reduced-form
coefficients are no longer proportional, so that the first moment of $T$ is
unrestricted. In this case, the instrumental variables estimand $\beta$ can be
interpreted as a weighted average of the marginal effect of the endogenous
variable on the outcome \citep{agi00}. I show that the unrestricted minimum
distance estimator coincides with a version of the bias-corrected two-stage
least squares estimator \citep{nagar59,dn01}, and use the \md\ objective
function to construct confidence intervals that remain valid when the
proportionality restriction is violated.

The \md\ objective function is also helpful in deriving a specification test
that is robust to many instruments. By testing the restriction on the first
moment of $T$, I derive a new test that is similar to that of \citet{cd93}, but
with an adjusted critical value. The adjustment ensures that the test is valid
under few strong as well as many instrument asymptotics that also allow for many
regressors. In contrast, when the number of regressors is allowed to increase
with the sample size, the size of the standard \citet{sargan58} specification
test converges to one, as does the size of the test proposed by \citet{ag11}.

The paper draws on two separate strands of literature. First, the literature on
many instruments that builds on the work by \citet{kunitomo80},
\citet{morimune83}, \citet{bekker94} and \citet{cs05}. Like \citet{anatolyev13},
I relax the assumption that the dimension of regressors is fixed, and I allow
them to grow with the sample size. \citet{hahn02}, \citet{chamberlain07},
\citet{cj09}, and \citet{moreira09as} focus on optimal inference with many
instruments when the errors are normal and homoscedastic, and my optimality
results build on theirs. Papers by \citet{hhn08}, \citet{akm10} and
\citet{vanhasselt10} relax the normality assumption. \citet{hnwcs12},
\citet{cshnw12}, \citet{chnsw14test} and \citet{BeCr15} also allow for
heteroscedasticity. An interesting new development is to employ shrinkage or
regularization to solve the incidental parameters problem (see, for example,
\citealp{bcch12}, \citealp{gt11}, or \citealp{carrasco12}). When combined with
additional assumptions on the model, these shrinkage estimators can be more
efficient than the efficient \md\ estimator proposed here.

Second, the literature on incidental parameters dating back to \citet{ns48}.
\citet{lancaster00} and \citet{arellano03} discuss the incidental parameters
problem in a panel data context. \citet{cm09ecta} relate invariance and random
effects approaches to the incidental parameters problem in a dynamic panel data
model. My results on the relationship between these two approaches in an
instrumental variables model build on theirs. \citet{sims00} proposes a similar
random-effects solution in a dynamic panel data model. \citet{moreira09as}
proposes to use invariance arguments to solve the incidental parameters problem.

The remainder of this paper is organized as follows. \Cref{sec:setup} sets up
the instrumental variables model, and reduces the data to the $T$ and $S$
statistics. \Cref{sec:random-effects} considers likelihood-based approaches to
inference under normality. \Cref{sec:minim-dist-interpr} relaxes the normality
assumption and considers the \md\ approach to inference. \Cref{sec:missp--clust}
considers \md\ estimation without imposing proportionality of the reduced-form
coefficients. \Cref{sec:specification-tests} studies tests of overidentifying
restrictions. \Cref{sec:conclusion} concludes. Proofs and derivations are
collected in the appendix. The supplemental appendix contains additional
derivations.

\section{Setup}\label{sec:setup}
In this section, I first introduce the model, notation, and the many instrument
asymptotic sequence that allows both the number of instruments and the number of
exogenous regressors to increase in proportion with the sample size. I then
reduce the data to the low-dimensional statistics $T$ and $S$, and define the
minimum distance objective function.

\subsection{Model and Assumptions}\label{sec:model-assumptions}
There is a sample of individuals $i=1,\dotsc, n$. For each individual, we observe
a scalar outcome $y_{i}$, a scalar endogenous regressor $x_{i}$,
$\Ne$-dimensional vector of exogenous regressors $w_{i}$, and $\Ni$-dimensional
vector of instruments $z_{i}^{*}$. The instruments and exogenous regressors are
treated as non-random.

It will be convenient to define the model in terms of an orthogonalized version
of the original instruments. To describe the orthogonalization, let $W$ denote
the $n\times \Ne$ matrix of regressors with $i$th row equal to $w_{i}'$, and let
$Z^{*}$ denote the $n\times \Ni$ matrix of instruments with $i$th row equal to
${z_{i}^{*}}'$. Let $\tilde{Z}=Z^{*}-W(W'W)^{-1}W'Z^{*}$ denote the residuals
from regressing $Z^{*}$ onto $W$. Then the orthogonalized instruments
$Z\in\mathbb{R}^{n\times \Ni}$ are given by $\Zp=\tilde{Z}R^{-1}$, where the
upper-triangular matrix $R\in\mathbb{R}^{\Ni\times \Ni}$ is the Cholesky factor
of $\tilde{Z}'\tilde{Z}$. Now, by construction, the columns of $Z$ are
orthogonal to each other as well as to the columns of $W$.\footnote{This
  orthogonalization is sometimes called a standardizing transformation; see
  \citet{phillips83} for discussion.}

Denote the $i$th row of $Z$ by $z_{i}'$, and let
$Y\in\mathbb{R}^{n\times 2}$ with rows $(y_{i}, x_{i})$ pool
all endogenous variables in the model. The reduced form regression of $Y$ onto
$\Zp$ and $W$ can be written as
\begin{align}\label{eq:reduced-form}
  Y&
     =\Zp\begin{pmatrix}
       \piss &\pifs
     \end{pmatrix}
               +W
              \begin{pmatrix}
                \pisse&\pifse
              \end{pmatrix}
                        +V,
\end{align}
where $V\in\mathbb{R}^{n\times 2}$ with rows $v_{i}'=(v_{1i}, v_{2i})$ pools the
reduced-form errors, which are assumed to be mean zero and homoscedastic,
\begin{align}\label{eq:homo}
  \E[v_{i}]&=0,&&\quad\text{and}& \E[v_{i}v_{i}']&=\Omega.
\end{align}
The reduced-form coefficients on the instruments are assumed to satisfy a
proportionality restriction, and the parameter of interest, $\beta$, corresponds
to the constant of proportionality:
\begin{assumptionc}{PR}[Proportionality restriction]\label{an:pr}
  $\piss=\pifs\beta$.
\end{assumptionc}
The proportionality restriction implies that
\begin{equation}\label{eq:structural-equation}
  y_{i}=x_{i}\beta+w_{i}'\beta^{w}_{n}+\epsilon_{i},
\end{equation}
where $\epsilon_{i}=v_{1i}-v_{2i}\beta$ is known as the structural error, and
$\beta^{w}_{n}=\pisse-\pifse\beta$. This equation is known as the structural
equation. In \Cref{sec:missp--clust}, I allow for certain violations of this
assumption, such as when the effect of $x_{i}$ on $y_{i}$ is heterogeneous.
Throughout the paper, I assume that $\Ni>1$, which implies that \Cref{an:pr} is
testable; I discuss specification testing in \Cref{sec:specification-tests}. In
order to employ sufficiency and invariance arguments, I assume that $v_{i}$ has
a normal distribution:
\begin{assumptionc}{N}[Normality]\label{an:n}
  The errors $v_{i}$ are i.i.d.\ and normally distributed.
\end{assumptionc}
This assumption has no effect on the consistency of estimators considered in
this paper, although it does affect their asymptotic distribution and asymptotic
efficiency. I drop this assumption in
\Cref{sec:minim-dist-interpr,sec:missp--clust,sec:specification-tests} when I
discuss a minimum-distance approach to inference.

To measure the strength of identification, I follow \citet{chamberlain07} and
\citet{ams08}, and use the parameter
\begin{align}\label{eq:lambdan}
  \lambdan&= \pifs'\pifs\cdot a'\Omega^{-1}a/n, & a&=
                                                    \begin{pmatrix}
                                                      \beta\\1
                                                    \end{pmatrix}.
\end{align}

The goal is to construct inference procedures that work well even if the number
of instruments $\Ni $ and the number of exogenous regressors $\Ne$ is large
relative to sample size. I therefore follow \citet{anatolyev13} and
\citet{kcfgi14} and allow both $\Ni $ and $\Ne $ to potentially grow with the
sample size:
\begin{assumptionc}{MI}[Many instruments]\label{an:mi}
  \begin{inparaenumi}
  \item\label{it:alphas} $\Ni /n=\alphai+o(n^{-1/2})$ and
    $\Ne /n=\alphae +o(n^{-1/2})$ for some $\alphae,\alphai\geq 0$ such that
    $\alphai+\alphae<1$;
  \item\label{it:mi:full-rank} The matrix $\bigl(W\;\; Z\bigr)$ is non-random
    and full column rank $\Ni+\Ne$;
  \item\label{it:z-lindeberg}
    $\sum_{i=1}^{n}((z_{i}'\pifs)^{4}+(z_{i}'\piss)^{4})/n^{2}=o(1)$; and
  \item\label{an:mi:lambda} $\lambdan= \lambda+o(1)$ for some $\lambda>0$.
  \end{inparaenumi}
\end{assumptionc}
\Cref{an:mi}~\ref{it:alphas} weakens the many instrument sequence of
\citet{bekker94} by allowing $\Ne $ to grow with the sample size. The motivation
for this is twofold. First, often the presence of many instruments is the result
of interacting a few basic instruments with many regressors \citep[as in, for
example][]{ak91}, in which case both $\Ne $ and $\Ni $ are large. Second,
oftentimes the instruments are valid only conditional on a large set of
regressors $w_{i}$; for example, if the instruments are randomly assigned within
a school, we need to condition on school fixed effects. By allowing
$\alphai=\alphae=0$, the assumption nests the standard few strong instrument
asymptotic sequence in which the number of instruments and regressors is fixed.
Parts~\ref{it:mi:full-rank}--\ref{an:mi:lambda} of \Cref{an:mi} are standard.
Part~\ref{it:mi:full-rank} is a normalization that requires excluding redundant
columns from $W$ and excluding columns of $Z$ that are redundant or already
included in $W$. It ensures that the reduced-form coefficients
in~\eqref{eq:reduced-form} are well-defined. Part~\ref{it:z-lindeberg} is used
to verify the Lindeberg condition. Part~\ref{an:mi:lambda} is the
many-instruments equivalent of the relevance assumption. It is equivalent to the
assumption that the concentration parameter \citep{rothenberg84}, given by
$\pifs '\pifs/\Omega_{22}$, grows at the same rate as the sample size.

\subsection{Sufficient statistics and limited information
  likelihood}\label{sec:suff-stat-limit}

Under normality, the set of sufficient statistics is given by the least-squares
estimators of the reduced-form coefficients $\Pin=\bigl(\piss\;\;\pifs\bigr)$
and $\Psin=\bigl(\pisse\;\;\pifse\bigr)$,
\begin{equation*}
  \begin{pmatrix}
    \hat\Pi\\\hat\Psi
  \end{pmatrix}
  =
  \begin{pmatrix}
    \Zp'{Y}\\
    (W'W)^{-1}W'{Y}
  \end{pmatrix}
  \in\mathbb{R}^{(\Ni +\Ne)\times 2},
\end{equation*}
and an unbiased estimator of the reduced-form covariance matrix $\Omega$ based
on the residual sum of squares,
\begin{equation}\label{eq:rss}
  S=\frac{1}{\nus} Y'(I_{n}-ZZ'-W'(W'W)^{-1}W)Y\in \mathbb{R}^{2\times 2},
\end{equation}
The advantage of working with the orthogonalized instruments is that now the
rows of $\hat{\Pi}$ are mutually independent. Since the distribution of
$\hat\Psi$ is unrestricted, we can drop it from the model and base inference on
$\hat\Pi$ and $S$ only as in \citet{moreira03} and \citet{ci04}. This step
eliminates the potentially high-dimensional nuisance parameter $\Psin$, so that
the model parameters are now given by the triplet $(\beta,\pifs,\Omega)$.

Estimators considered in this paper will only depend on $\hat{\Pi}$ through the
statistic
\begin{equation}\label{eq:T}
  T=\frac{1}{n}{\hat\Pi}' \hat\Pi=
\frac{1}{n}Y'ZZ'Y
\in\mathbb{R}^{2\times 2}.
\end{equation}
Define the following functions of the statistics $T$ and $S$:
\begin{align*}
  \Qs(\beta,\Omega)&=\frac{b'T b}{b'\Omega b},& \Qt(\beta,\Omega)&=
                                                                   \frac{a'\Omega^{-1}T \Omega^{-1}a}{a'\Omega^{-1} a},& b&=
                                                                                                                            \begin{pmatrix}
                                                                                                                              1\\-\beta
                                                                                                                            \end{pmatrix},
\end{align*}
and let $m_{\min}$ and $m_{\max}$ denote the minimum and maximum eigenvalues of
the matrix $S^{-1}T$.

The likelihood of the model~\eqref{eq:reduced-form}--\eqref{eq:homo} under
\Cref{an:pr,an:n} is known as the limited information likelihood \citep{ar49}.
The limited information maximum likelihood (\liml) estimator of $\beta$ is given
by
\begin{align}\label{eq:hbetaliml}
  \h{\beta}{liml}& =\argmax_{\beta} \Qt(\beta,S)=
                   \argmin_{\beta}\Qs(\beta,S)=\frac{T_{12}-m_{\min}S_{12}}{T_{22}-m_{\min}S_{22}}.
\end{align}
It turns out that $\h{\beta}{liml}$ is consistent and asymptotically normal
under \Cref{an:mi} despite the incidental parameters problem \citep{bekker94}. I
will give some insight into this result in \Cref{sec:random-effects}.

Due to the incidental parameters problem, the $(\beta,\beta)$ block of the
inverse information matrix of the limited information likelihood does not yield
a consistent estimate of the asymptotic variance of \liml\@. The asymptotic
distribution of $\h{\beta}{liml}$ under \Cref{an:pr,,an:mi,,an:n} is given by
(see \citealp{bekker94} and \citealp{kcfgi14} for derivation)
\begin{equation}\label{eq:avar:liml:formula}
  \sqrt{n}\left(\h{\beta}{liml}-\beta\right)\indist
  \mathcal{N}\left(0,\mathcal{V}_{\liml,N}\right),
\end{equation}
where $\indist$ denotes convergence in distribution, and
\begin{equation}\label{eq:avar:liml}
  \mathcal{V}_{\liml,N}=  \frac{b'\Omega b\cdot a'\Omega^{-1}a}{\lambda}\left(
    1+\frac{\alphai(1-\alphae)}{1-\alphai-\alphae}
    \frac{1}{\lambda}\right).
\end{equation}
In contrast, the $(\beta,\beta)$ block of the inverse information matrix is
given by $b'\Omega b\cdot a'\Omega^{-1}a/(n\lambdan)$, missing the correction
factor in parentheses (see supplemental appendix for derivation). This
correction factor can be substantial even when the ratio of instruments to
sample size, $\alphai$, is small if $\lambda$ is small.

\subsection{Using invariance to reduce the dimension of the parameter space}%
\label{sec:using-invar-reduce}

To reduce the dimension of the parameter space, I follow \citet{ams06},
\citet{chamberlain07}, \citet{cj09}, and \citet{moreira09as}, and require
decision rules (procedures used for constructing point estimates and confidence
intervals from the data) to be invariant with respect to rotations of the
instruments. In other words, changing the co-ordinate system for the instruments
should not affect inference about $\beta$---if we re-order the instruments, or
use a different orthogonalization procedure to construct $\Zp$, we should get
the same point estimate and confidence interval for $\beta$. A decision rule is
invariant under rotations of instruments if it remains unchanged under the
transformation $(\hat{\Pi},S)\mapsto(g\hat{\Pi},S)$, where
$g\in\mathcal{O}(\Ni)$, the group of $\Ni\times \Ni$ orthogonal matrices. A
necessary and sufficient condition for a decision rule to be invariant is that
it depends on the data only through a maximal invariant \citep[Theorem
2.3]{eaton89}. A statistic $m(\hat{\Pi},S)$ is maximal invariant if (i)
$m(\hat{\Pi},S)=m(g\hat{\Pi},S)$ for all $g\in\mathcal{O}(\Ni)$; and (ii)
whenever $m(\dot{\Pi},\dot{S})=m(\ddot{\Pi},\ddot{S})$ for some
$\dot{\Pi},\ddot{\Pi}\in\mathbb{R}^{\Ni\times 2}$ and
$\dot{S},\ddot{S}\in\mathbb{R}^{2\times 2}$, then
$(\dot{\Pi},\dot{S})=(g\ddot{\Pi},\ddot{S})$ for some $g\in\mathcal{O}(\Ni)$. It
is straightforward to check that the pair of matrices $(S,T)$ is a maximal
invariant statistic. The distribution of $(S,T)$ depends on $\pifs$ only through
$\pifs'\pifs$, or equivalently through
$\lambdan=\pifs'\pifs\cdot a'\Omega^{-1}a/n$. This reduces the parameter space
to $(\beta,\lambdan,\Omega)$, which has a fixed
dimension.\footnote{\label{fn:more-than-one-endog}Similar arguments can be used
  to generalize the results in this paper to the case with more than one
  endogenous variable. In particular, if $\dim(x_{i})=J$ and
  $\dim(Y)=n\times (J+1)$, then one can use invariance arguments to reduce the
  data to the same pair of matrices $(S,T)$ defined in~\eqref{eq:rss}
  and~\eqref{eq:T}, now with dimension $(J+1)\times (J+1)$.}

There are two general approaches to constructing invariant decision rules based
on the maximal invariant $(S,T)$. First, one can use the likelihood based on $S$
and $T$, called the invariant likelihood, $\linv(\beta,\lambdan,\Omega;S,T)$. I
consider this approach in detail in \Cref{sec:random-effects}. The disadvantage
of this approach is that the validity of inference based on the invariant
likelihood is sensitive to \Cref{an:n}.

The second approach is to use moment restrictions on $S$ and $T$ implied by the
model. In particular, the reduced form~\eqref{eq:reduced-form}--\eqref{eq:homo}
without any further assumptions implies
\begin{subequations}\label{eq:md:moment-conditions}
  \begin{align}
    \E[S]
    &=\Omega,\label{eq:md:moment-conditions:S}\\
    \E[T-(\Ni/n) S]
    &=\Xi_{n},
    & \text{where}\quad\Xi_{n}
    &= \frac{1}{n}
      \begin{pmatrix}
        {\piss}&    \pifs\\
      \end{pmatrix}'\begin{pmatrix}
        {\piss}&    \pifs\\
      \end{pmatrix}.\label{eq:md:moment-conditions:S-T}
  \end{align}
\end{subequations}
Under \Cref{an:pr}, the matrix of second moments of the reduced-form
coefficients, $\Xi_{n}$, has reduced rank,
\begin{equation}\label{eq:rank-restriction}
  \Xi_{n}=\Xi_{22,n} aa'=\Xi_{22,n}
  \begin{pmatrix}
    \beta^{2}& \beta\\
    \beta& 1
  \end{pmatrix},
\end{equation}
with $\Xi_{22,n}=\pifs'\pifs/n=\lambdan/(a'\Omega^{-1}a)$. This rank restriction
can be used to build a minimum distance objective
function\footnote{\label{fn:vech}The operator $\vech(A)$ transforms the
  lower-triangular part of $A$ into a single column---when $A$ is symmetric, as
  is the case here, the operator can be thought of as vectorizing $A$ while
  removing the duplicates.}
\begin{equation}\label{eq:md:objective2}
  \mathcal{Q}_{n}(\beta,\Xi_{22,n};\hat W_{n})=\vech\left(\textstyle
    T-(\Ni/n)S-\Xi_{22,n} aa'\right)'\hat W_{n}\vech\left(\textstyle
    T-(\Ni/n)S-\Xi_{22,n}aa'\right),
\end{equation}
where $\hat{W}_{n}\in\mathbb{R}^{3\times 3}$ is some weight matrix. Since the
nuisance parameter $\Omega$ only appears in the moment
condition~\eqref{eq:md:moment-conditions:S}, which is unrestricted, we can
exclude the moment condition from the objective
function~\eqref{eq:md:objective2} without any loss of information \citep[Section
3.2]{chamberlain82joe}. I consider this approach in detail in
\Cref{sec:minim-dist-interpr,sec:missp--clust,sec:specification-tests}, where I
show that this approach is more attractive once \Cref{an:n} is relaxed.

\section{Likelihood-based estimation and inference}\label{sec:random-effects}

This section shows that by combining the invariant likelihood with a particular
prior on $\lambdan$, we can construct a likelihood with a simple closed form
that addresses the incidental parameters problem. I show that this likelihood is
equivalent to the random effects likelihood of \citet{ci04}, and that maximizing
it yields the \liml\ estimator of $\beta$.

First consider maximizing the invariant likelihood. To state the result, let
$\omegan=\pifs/\norm{\pifs}$, so that $\hat{\Pi}$ and $S$ can be parametrized by
$(\beta,\omegan,\lambdan,\Omega)$. The parameter $\omegan$ lies on the unit
sphere $\mathbb{S}^{\Ni-1}$ in $\mathbb{R}^{\Ni}$ and it can be thought of as
measuring the relative strength of the individual instruments.
\begin{lemma}\label{th:invariant-lik}
  The invariant likelihood
  $\mathcal{L}_{\text{\an{inv}},n}(\beta,\lambdan,\Omega;S,T)$ is maximized over
  $\beta$ at $\h{\beta}{liml}$. This result also holds if $\lambdan$ is fixed at
  an arbitrary value. Furthermore,
  \begin{equation}\label{eq:invariant-lik-integrated}
    \mathcal{L}_{\text{\an{inv}},n}(\beta,\lambdan,\Omega;S,T)=\int_{\mathbb{S}^{\Ni -1}}
    \mathcal{L}_{\li,n}(\beta,\lambdan,\omegan, \Omega;\hat\Pi,S)\,\dd
    F_{\omegan}(\omegan),
  \end{equation}
  where $\mathcal{L}_{\li,n}$ denotes the limited information likelihood, and
  $F_{\omegan}(\cdot)$ denotes the uniform distribution on the unit sphere
  $\mathbb{S}^{\Ni-1}$.
\end{lemma}
The first part of \Cref{th:invariant-lik} generalizes the result in
\citet{moreira09as} that the maximum invariant likelihood estimator for $\beta$
coincides with \an{limlk} when $\Omega$ is known. It also explains why the
limited information likelihood produces an estimator that is robust to many
instruments, even though the number of parameters in the likelihood increases
with sample size: it is because \an{liml} happens to coincide with the maximum
invariant likelihood estimator.

The last part of \Cref{th:invariant-lik} shows that the invariant likelihood is
equivalent to the integrated (marginal) likelihood that puts a uniform prior on
$\omegan$. This observation will allow me to build the connection between the
invariant likelihood and the random-effects likelihood of \citet{ci04}. In
particular, consider integrating the limited information likelihood with respect
to the following prior on $\lambdan$, in addition to the uniform prior on
$\omegan$:
\begin{equation}\label{eq:chi-square-prior}
  \lambdan\sim \frac{\lambda}{\Ni}\chi^{2}(\Ni).
\end{equation}
The hyperparameter $\lambda$ corresponds to the limit of $\lambdan$ under
\Cref{an:mi}. I allow it to be determined by the data, so that the prior will be
dominated in large samples. The two priors on $\omegan$ and $\lambdan$ are
equivalent to a single normal prior over the scaled first-stage coefficients
$\etan=\pifs \sqrt{a'\Omega^{-1}a/n}$,
\begin{equation}\label{eq:re}
  \eta_{n}\sim \mathcal{N}(0,\lambda/\Ni\cdot I_{\Ni}).
\end{equation}
This prior is the random-effects prior proposed in \citet{ci04}. Therefore, the
integrated likelihood obtained after integrating the limited information
likelihood over the uniform prior on $\omega_{n}$ and the chi-square prior on
$\lambdan$ coincides with the \re\ likelihood that integrates the limited
information likelihood over the normal prior~\eqref{eq:re}. The \re\ likelihood
has a simple closed form:\footnote{\citet{ci04} also consider putting a random
  effects prior only on some coefficients; the coefficients on the remaining
  instruments are then assumed to be fixed. When referring to the random-effects
  likelihood, I assume that we put a random-effects prior on all coefficients.}
\begin{equation}\label{eq:re:lik}
  \begin{split}
    \mathcal{L}_{\re,n}(\beta, \lambda, \Omega)&= \int_{\mathbb{R}^{\Ni}}
    \mathcal{L}_{\text{\an{li}},n}(\beta,\etan,\omega_{n},\Omega;\hat{\Pi},S)\,\dd
    F_{\etan\mid \lambda}(\etan\mid \lambda) \\
    &= \int_{\mathbb{R}}\int_{\mathbb{S}^{\Ni -1}}
    \mathcal{L}_{\li,n}(\beta,\lambdan,\omega_{n},\Omega;\hat\Pi,S)\,\dd
    F_{\omega_{n}}(\omega_{n})\,\dd F_{\lambdan\mid \lambda}(\lambdan\mid
    \lambda)
    \\
    & =
    \abs{S}^{\frac{\nus-3}{2}}\cdot \left(1 +
      \frac{n}{\Ni }\lambda\right)^{ - \Ni /2} \abs{\Omega}^{-\frac{n-\Ne}{2}}
    e^{-\frac{1}{2}\left(\trace(\Omega^{ - 1}\tilde{S}) -
        \frac{n\lambda\Qt(\beta, \Omega)}{\Ni/n + \lambda}\right)},
  \end{split}
\end{equation}
where the last equality holds up to a normalizing constant, and
$\tilde{S}=(\nus)S+nT$. \citet{ci04} motivate the \re\ prior as a modeling tool:
since the prior has zero mean, it intuitively captures the idea that the
individual instruments may not be very relevant. This motivation leaves it
unclear however, whether inference based on the \re\ is asymptotically valid
when the first-stage parameters are viewed as fixed. The
equivalence~\eqref{eq:re:lik} implies that one can indeed use the \re\
likelihood for inference. In particular, since the invariant likelihood has a
fixed number of parameters and the invariant model is locally asymptotically
normal \citep{cj09}, inference based on it will be asymptotically valid by
standard arguments. Since the prior on $\lambdan$ gets dominated in large
samples, this implies that inference based on the \re\ likelihood will also be
asymptotically valid. Furthermore, since by~\Cref{th:invariant-lik} constraining
$\lambdan$ does not affect the maximum invariant likelihood estimator for
$\beta$, integrating the invariant likelihood with respect to the chi-square
prior~\eqref{eq:chi-square-prior} will not affect it either: it will still be
given by $\h{\beta}{liml}$. The next proposition summarizes and formalizes these
results.
\begin{proposition}\label{th:re-theorem}
  \mbox{}
  \begin{compactenumi}
  \item\label{it:th:re:max:ml} The \re\ likelihood~\eqref{eq:re:lik} is
    maximized at
    \begin{align*}
      \h{\beta}{re}&=\h{\beta}{liml},\\
      \h{\lambda}{re}&=\max\{m_{\max}-\Ni /n,0\},\\
      \h{\Omega}{re}&=\frac{\nus}{n-\Ne}S+\frac{n}{n-\Ne}\left(T-
                      \frac{\h{\lambda}{re}}{\h{a}{re}'S^{-1}\h{a}{re} }
                      \h{a}{re}\h{a}{re}'\right),\qquad \h{a}{re}=
                      \Bigl(\begin{smallmatrix}
                        \h{\beta}{re}\\1
                      \end{smallmatrix}\Bigr).
    \end{align*}
  \item\label{it:re-hessian} If $m_{\max}>\Ni/n$, the (1,1) element of the
    inverse Hessian of the \re\ likelihood~\eqref{eq:re:lik}, evaluated at
    $(\h{\beta}{re},\h{\lambda}{re},\h{\Omega}{re})$, is given by:
    \begin{equation*}
      \h{\mathcal{H}}{re}^{11}= \frac{\h{b}{re}'\h{\Omega}{re}\h{b}{re}
        (\h{\lambda}{re}+\Ni/n)}{n\h{\lambda}{re}}\left(
        \hQs\h{\Omega}{re,22}-T_{22}+\frac{\hat c}{1-\hat c}\frac{\hQs}{\h{a}{re}'
          \h{\Omega}{re}^{-1}\h{a}{re}} \right)^{-1},
    \end{equation*}
    where $\hQs= \Qs(\h{\beta}{re},\h{\Omega}{re})$,
    $\hat{c}=\frac{\h{\lambda}{re} \hQs}{(\Ni/n+\h{\lambda}{re})(1-\Ne/n)}$, and
    $\h{b}{re}=(1,-\h{\beta}{re})'$.
  \item\label{it:re-hessian-cons} Under \Cref{an:pr,,an:n,an:mi},
    $-n\h{\mathcal{H}}{re}^{11}\inprob \mathcal{V}_{\liml,N}$, with
    $\mathcal{V}_{\liml,N}$ given in Equation~\eqref{eq:avar:liml}.
  \end{compactenumi}
\end{proposition}
Part~\ref{it:th:re:max:ml} of \Cref{th:re-theorem} formalizes the claim that the
estimator of $\beta$ remains unchanged under the additional chi-square prior for
$\lambdan$. Part~\ref{it:re-hessian} derives the expression for the inverse
Hessian. The condition $m_{\max}> \Ni/n$ makes sure that the constraint
$\lambda\geq 0$ does not bind, otherwise the Hessian is singular. It holds with
probability approaching one under Assumption~\ref{an:mi}~\ref{an:mi:lambda}, as
$m_{\max}-\Ni/n\inprob \lambda>0$. Part~\ref{it:re-hessian-cons} proves that the
extra prior on $\lambdan$ gets dominated in large samples so that the inverse
Hessian can be used to estimate the asymptotic variance of $\h{\beta}{liml}$
(one could also use the inverse Hessian of the invariant likelihood, although
this involves numerical optimization since maximum invariant likelihood
estimates of $\lambdan$ and $\Omega$ are not available in closed form). It is
important that the prior on $\lambdan$ is chosen such that the prior is
dominated in large samples. For example, \citet{lancaster02} suggests
integrating the orthogonalized incidental parameters out with respect to a
uniform prior. Here such prior corresponds to a flat prior on $\etan$, which is
equivalent to a uniform prior on $\omega_{n}$, and an improper prior on
$\lambdan$, obtained by taking the limit of~\eqref{eq:chi-square-prior} as
$\lambda\to \infty$. However, this improper prior on $\lambdan$ will never get
dominated by the data, and as a result, it can be shown that the resulting
likelihood will fail to produce valid confidence intervals.

\section{Minimum distance estimation and
  inference}\label{sec:minim-dist-interpr}

In this section, I first show that the random effects estimator is in fact
equivalent to a minimum distance estimator that uses a particular weight matrix.
This weight matrix weights the restrictions efficiently under normality, but not
otherwise. I derive a new estimator of $\beta$ based on the efficient weight
matrix that is more efficient than \liml\ when the normality assumption is
dropped. Moreover, unlike inference based on the random effects likelihood,
minimum-distance-based inference will be asymptotically valid even if the
reduced-form errors are not normally distributed.

To simplify the expressions in this section, let
$D_{d}\in\mathbb{R}^{d^{2}\times d(d+1)/2},L_{d}\in\mathbb{R}^{d(d+1)/2\times
  d^{2}}$, and $N_{d}\in\mathbb{R}^{d^{2}\times d^{2}}$ denote the duplication
matrix, the elimination matrix, and the symmetrizer matrix, respectively (see
\citet{mn80} for definitions of these matrices). The symmetrizer matrix has the
property that for any $d\times d$ matrix $A$,
$N_{d}\mkvec(A)=(1/2)\mkvec(A+A')$. The duplication matrix transforms the
$\vech$ operator into a $\mkvec$ operator, and the elimination operator performs
the reverse operation, so that for a symmetric $d\times d$ matrix $A$,
$D_{d}\vech(A)=\mkvec(A)$, and $L_{d}\mkvec(A)=\vech(A)$.

\subsection{Random effects and minimum distance}\label{sec:rand-effects-minim}
The random effects likelihood~\eqref{eq:re:lik} and the minimum distance
objective function~\eqref{eq:md:objective2} both leverage the rank
restriction~\eqref{eq:rank-restriction} to construct an estimator of $\beta$.
There should therefore exist a weight matrix such that the random effects
estimator of $(\beta,\Xi_{22,n})$ is asymptotically equivalent to a minimum
distance estimator with respect to this weight matrix. The next proposition
shows that if the weight matrix is appropriately chosen, the minimum distance
and random effects estimators are in fact \emph{identical}.
\begin{proposition}\label{th:md-re:equivalence2}
  Consider the minimum distance objective function~\eqref{eq:md:objective2} with
  respect to the weight matrix $\h{W}{re}=D_{2}'(S^{-1}\otimes S^{-1})D_{2}$.
  \begin{inparaenumi}
  \item\label{item:md-re-1} The objective function is minimized at
    $(\h{\beta}{re},\h{\Xi}{22,re})$, where
    $\h{\Xi}{22,re}=\h{\lambda}{re}/(\h{a}{re}'\h{\Omega}{re}^{-1}\h{a}{re})$
  \item Under \Cref{an:pr,,an:n,an:mi}, the weight matrix $\h{W}{re}$ is
    asymptotically optimal.
  \end{inparaenumi}
\end{proposition}
The second part of \Cref{th:md-re:equivalence2} shows that if the errors are
normally distributed, then the random effects weight matrix $\h{W}{re}$ weights
the moment condition~\eqref{eq:md:moment-conditions:S-T} efficiently under
many-instrument asymptotics, even though $\h{W}{re}$ doesn't converge to the
inverse of the asymptotic variance of the moment condition. The proof shows that
the inverse of the asymptotic variance is not the unique optimal weight matrix,
but that there exists a whole class of optimal weight matrices, and that this
class includes $\h{W}{re}$.\footnote{ The standard condition that the weight
  matrix converges to the inverse of the asymptotic covariance matrix of the
  moment conditions is sufficient, but not necessary for asymptotic efficiency
  \citep[Section 5.2]{nm94}.} As I show in the next subsection, this optimality
result is sensitive to \Cref{an:n}.

The equivalence between minimum distance and \re\ estimators is related to the
observation in \citet{bekker94} that \liml\ can be thought of as a
method-of-moments estimator in the sense that it satisfies
$(T-m_{\min}S)(1,-\h{\beta}{liml})'=0$, which is similar to a first-order
condition of the objective function~\eqref{eq:md:objective2} when the weight
$\hat{W}_{\re}$ is used. It is also related to \citet{go71}, who consider a
minimum distance objective function based on the proportionality
restriction~\ref{an:pr},
\begin{equation}\label{eq:md:go}
  \mathcal{Q}_{\text{\an{go}},n}(\beta,\pifs)=
  \mkvec\left(\hat\Pi-\pifs
    a'\right)'\left(S^{-1}\otimes
    I_{\Ni}\right)\mkvec\left(\hat\Pi-\pifs a'\right).
\end{equation}
\citet{go71} show that this objective function is minimized at
$\h{\beta}{liml}$. However, the number of parameters in this objective function
diverges to infinity under \Cref{an:mi}, so it cannot be used for inference.

\subsection{Minimum distance estimation under non-normal
  errors}\label{sec:effic-minim-dist}

The efficiency of $\h{\beta}{liml}$ as well as the expression for the asymptotic
distribution of $\h{\beta}{liml}$ given in~\eqref{eq:avar:liml} depend on
\Cref{an:n}. This sensitivity to the normality assumption is similar to the
efficiency results for the maximum likelihood estimator in panel-data models in
which identification is based on covariance restrictions \citep[Chapter
5.4]{arellano03}.

In order to derive the optimal weight matrix as well as the correct asymptotic
variance formulae under non-normality, we first need the limiting distribution
of the moment condition~\eqref{eq:md:moment-conditions:S-T}. The moment
condition depends on the data through the three-dimensional statistic
$\vech(T-(\Ni/n)S)$. It can be seen from the definition of $T$ and $S$ given
in~\Cref{eq:rss,,eq:T} that this statistic can be written as a quadratic form
\begin{equation*}
  T-\frac{\Ni}{n}S=\frac{1}{n}Y'HY=\frac{1}{n}(\Zp\pifs a'+V)'H(\Zp\pifs a'+V),
\end{equation*}
where
\begin{equation*}
  H=\Zp\Zp'-\frac{\Ni}{\nus}(I_{n}-W(W'W)^{-1}W'-\Zp\Zp').
\end{equation*}

We need to impose some regularity conditions on the components of the quadratic
form. Let $\diag(A)$ denote the $n$-vector consisting of diagonal elements of an
$n$-by-$n$ matrix $A$, and let $\delta_{n}=\diag(H)'\diag(H)/\Ni$.
\begin{assumptionc}{RC}[Regularity conditions]\label{an:rc}
  \begin{inparaenumi}
  \item\label{it:8-moments} The errors $v_{i}$ are i.i.d, with finite $8$th
    moments;
  \item\label{it:d-limits} For some $\delta,\mu\in\mathbb{R}$, as $n\to\infty$,
    $\delta_{n}\to \delta$, and
    $\pifs'\Zp'\diag(H)/\sqrt{n\Ni}\to\mu$
  \end{inparaenumi}
\end{assumptionc}
Part~\ref{it:8-moments} relaxes the normality assumption on the errors.
Part~\ref{it:d-limits} ensures the asymptotic covariance matrix is well-defined.

\begin{lemma}\label{th:moment-variance}
  Under \Cref{an:pr,,an:mi,an:rc}:
  \begin{compactenumi}
  \item\label{it:moment-variance}
    $\sqrt{n}\vech( T-(\Ni/n)S- \Xi_{22,n}aa')\indist \mathcal{N}(0, \Delta)$,
    with $\Delta=L_{2}(\Delta_{1}+\Delta_{2}+\Delta_{3}+\Delta_{3}')L_{2}'$,
    where
    \begin{align*}
      \Delta_{1}
      &=2N_{2}\left(\Xi_{22}aa'\otimes \Omega + \Omega\otimes
        \Xi_{22}aa'+\tau \Omega\otimes \Omega,\right),
      & \tau&=\alphai(1-\alphae)/(\alphanu),\\
      \Delta_{2}
      &=\alphai\delta \left[\Psi_{4}-\mkvec(\Omega)
        \mkvec(\Omega)'-2N_{2}( \Omega\otimes \Omega)\right],
      & \Psi_{4}&=\E[(v_{i}v_{i}')\otimes(v_{i}v_{i}')],\\
      \Delta_{3}&=2N_{2}(\sqrt{\alphai}\mu \Psi_{3}'\otimes a),
      &\Psi_{3}&=\E[(v_{i}v_{i}')\otimes v_{i}],
    \end{align*}
    and $\Xi_{22}=\lambda/(a'\Omega^{-1}a)$.
  \item\label{it:variance-estimator} Let $M=I_{n}-\Zp\Zp'-W(W'W)^{-1}W'$, let
    $\hat{V}=MY$ with rows $\hat{v}_{i}'$ denote estimates of the reduced form
    errors, and let $\hat{\pi}_{2}$ denote the second column of $\hat{\Pi}$.
    Then
    \begin{align*}
      \hat\Psi_{3}
      & =\frac{\sum_{i}[(\hat v_{i} \hat v_{i}')\otimes
        \hat{v}_{i}]}{\sum_{i,j}M_{ij}^{3}} \inprob \Psi_{3},\\
      \hat\Psi_{4}
      &=\frac{ \sum_{i}(\hat v_{i} \hat v'_{i})\otimes (\hat v_{i}
        \hat v_{i}')-
        \left[\sum_{i}M_{ii}^{2}-\sum_{i,j}M_{ij}^{4}\right](2N_{2}
        \hat\Omega\otimes \hat\Omega+\mkvec(\hat\Omega)\mkvec(\hat\Omega)')
        }{\sum_{i,j} M_{ij}^{4}}\inprob \Psi_{4},
    \end{align*}
    and $\hat{\mu}=\hat{\pi}_{2}Z'\diag(H)/\sqrt{n\Ni}\inprob\mu$.
  \end{compactenumi}
\end{lemma}
Part~\ref{it:moment-variance} shows that the asymptotic variance consists of
three distinct terms. If the errors are normally distributed, then
$\Delta_{2}=\Delta_{3}=0$. The term $\Delta_{2}$ accounts for excess kurtosis of
the errors, and the term $\Delta_{3}$ accounts for skewness.
Part~\ref{it:variance-estimator} provides consistent estimators for the third
and fourth moments of the errors, and for $\mu$. Since the probability limits of
$S$ and $T$ do not depend on \Cref{an:n}, the other components of
$\Delta_{1},\Delta_{2}$ and $\Delta_{3}$ can be consistently estimated by
$\h{\beta}{re}$, $\h{\Omega}{re}$, and
$\h{\Xi}{22,re}=\h{\lambda}{re}/(\h{a}{re}'\h{\Omega}{re}^{-1}\h{a}{re})$.
Therefore, a consistent estimator of the asymptotic covariance matrix $\Delta$
is given by
\begin{equation}\label{eq:hatdelta}
  \hat\Delta=L_{2}(\hat\Delta_{1}+\hat\Delta_{2}+\hat\Delta_{3}
  +\hat\Delta_{3}')L_{2}',
\end{equation}
where the terms $\hat{\Delta}_{1}$, $\hat{\Delta}_{2}$, and $\hat{\Delta}_{3}$
are given by replacing $\beta$, $\Xi_{22}$, and $\Omega$ in the definitions of
$\Delta_{1},\Delta_{2}$, and $\Delta_{3}$ by their random-effects estimators,
replacing $\Psi_{3}$ and $\Psi_{4}$ by $\hat\Psi_{3}$ and $\hat\Psi_{4}$, and
replacing $\delta$ and $\mu$ by $\delta_{n}$ and $\hat{\mu}$.

\paragraph{Inference based on LIML}
Since $\h{\beta}{liml}$ is a minimum distance estimator, its asymptotic variance
is given by the (1,1) element of the matrix
\begin{equation}\label{eq:gmm:avar}
  (G'WG)^{-1}G'W\Delta WG(G'WG)^{-1},
\end{equation}
where $W=D_{2}'(\Omega^{-1}\otimes \Omega^{-1})D_{2}=\plim \h{W}{re}$, and $G$
is the derivative of the moment condition~\eqref{eq:md:moment-conditions:S-T},
\begin{equation*}
  G=L_{2}
  \begin{pmatrix}
    \Xi_{22} \left(a\otimes \bigl(\begin{smallmatrix} 1\\0
      \end{smallmatrix}\bigr)+ \bigl(\begin{smallmatrix} 1\\0
      \end{smallmatrix}\bigr)\otimes a\right)& a\otimes a
  \end{pmatrix}.
\end{equation*}
This element evaluates as
\begin{equation*}
  \mathcal{V}_{\liml}=
  \mathcal{V}_{\liml,N}+
  \frac{2\sqrt{\alphai}\mu}{\Xi_{22}^{2}}\E[(v_{2i}-\gamma\epsilon_{i})
  \epsilon_{i}^{2}]+
  \frac{\alphai\delta}{\Xi_{22}^{2}}
  \E[\epsilon_{i}^{2}(v_{2i}-\gamma\epsilon_{i})^{2}-\abs{\Omega}],
\end{equation*}
where $\epsilon_{i}=v_{1i}-v_{2i}\beta$ is the structural error, and $\gamma$ is
regression coefficient from projecting $v_{2i}$ onto it,
\begin{equation}\label{eq:gamma-defn}
  \gamma=(\Omega_{12}-\Omega_{22}\beta)/ (b'\Omega b),
\end{equation}
The term $\mathcal{V}_{\liml,N}$ (given in Equation~\eqref{eq:avar:liml})
corresponds to the asymptotic variance of $\h{\beta}{liml}$ under normal errors.
The two remaining terms are corrections for skewness and excess kurtosis.
\citet{anatolyev13} derives the same asymptotic variance expression by working
with the explicit definition of $\h{\beta}{liml}$. If $\alphae=0$, then
$\mathcal{V}_{\liml}$ reduces to the asymptotic variance given in \citet{hhn08},
\citet{akm10}, and \citet{vanhasselt10}. Due to the presence of the two extra
terms, the inverse Hessian will no longer estimate the asymptotic variance
consistently. However, a consistent plug-in estimator of~\eqref{eq:gmm:avar} can
easily be computed by replacing $\Delta$ by $\hat\Delta$ and replacing $a$,
$\Xi_{22}$, and $\Omega$ in the expressions for $G$ and $W$ by $\h{a}{re}$,
$\h{\Xi}{22,re}$ and $\h{\Omega}{re}$.

\paragraph{Efficient minimum distance estimator}
Using the inverse of the variance estimator~\eqref{eq:hatdelta} as a weight
matrix in the minimum distance objective function yields an efficient minimum
distance (\emd) estimator
\begin{equation*}
  (\h{\beta}{emd},\h{\Xi}{22,emd})=
  \argmin_{\beta,\Xi_{22}}
  \mathcal{Q}_{n}(\beta,\Xi_{22,n};\hat\Delta^{-1}).
\end{equation*}
Since the objective function is a fourth-order polynomial in two arguments, the
solution can be easily found numerically. It then follows by standard arguments
(see, for example, \citet{nm94}), that when $\alphai>0$,
\begin{equation*}
  \sqrt{n}(\h{\beta}{emd}-\beta)\indist \mathcal{N}(0,\mathcal{V}_{\emd}),
\end{equation*}
where $\mathcal{V}_{\emd}$ corresponds to (1,1) element of the matrix
$(G'\Delta^{-1} G)^{-1}$, which evaluates as
\begin{equation}\label{eq:vemd}
  \mathcal{V}_{\emd}=  \mathcal{V}_{\liml} -
  \frac{1}{\Xi_{22}^{2}(b'\Omega b)^{2}}\frac{ \left(
      \sqrt{\alphai}\mu\E[\epsilon_{i}^{3}]
      +\alphai\delta\E[(v_{2i}-\gamma\epsilon_{i})\epsilon_{i}^{3}]\right)^{2}
  }{2\tau+\alphai\delta \kappa},
\end{equation}
where
\begin{equation}
  \label{eq:ek}
  \kappa=  \E[\epsilon_{i}^{4}/(b'\Omega b)^{2}-3]
\end{equation}
measures excess kurtosis of $\epsilon_{i}$. A consistent plug-in estimator of
$\mathcal{V}_{\emd}$ can be easily constructed by replacing $\Delta$ by
$\hat\Delta$, and replacing $\Xi_{22}$ and $\beta$ in the expression for $G$ by
their random-effects, or \emd\ estimators.

There is a slightly stronger sense in which $\h{\beta}{emd}$ is efficient than
just being efficient in the class of minimum distance estimators: it exhausts
the information available in $(S,T)$. In particular, as argued in
\citet{vdPBe95}, the efficiency bound for estimators that are smooth functions
of $(S,T)$ is given by the efficient minimum distance estimator based on the
moment conditions~\eqref{eq:md:moment-conditions:S}
and~\eqref{eq:md:moment-conditions:S-T}. However, since the nuisance parameter
$\Omega$ only appears in the first moment
condition~\eqref{eq:md:moment-conditions:S}, which is unrestricted, we can
exclude it from the objective function, and the minimum distance estimator of
$\beta$ with respect to an efficient weight matrix will achieve the same
asymptotic variance \citep[Section 3.2]{chamberlain82joe}.

\citet{hahn02} shows that when the errors are restricted to be normal, an
estimator that exhausts the information in $(S,T)$ will have variance given by
$\mathcal{V}_{\liml}$. \citet{akm10} generalize this result by allowing the
errors to belong to the family of elliptically contoured
distributions.\footnote{\label{fn:elliptical-dist}A mean-zero random vector has
  an elliptically contoured distribution if its characteristic function can be
  written as $\varphi(t'\mathcal{V} t)$, for some matrix $\mathcal{V}$ and some
  function $\varphi$. The multivariate normal distribution is a special case,
  with $\varphi(t)=e^{-t't/2}$.} Equation~\eqref{eq:vemd} shows that this is not
true in general. Indeed, the \citet{akm10} result obtains as a special case,
since for elliptically contoured distributions, $\Psi_{3}=0$, so that
$\E[\epsilon_{i}^{3}]=0$ and $\Psi_{4}$ is proportional to
$\mkvec(\Omega) \mkvec(\Omega)'+2N_{2}\Omega\otimes \Omega$ \citep{ww92}, which
implies
$\E[(v_{2i}-\gamma\epsilon_{i})\epsilon_{i}^{3}]
=0$, so that the second term in~\eqref{eq:vemd}---the efficiency gain over
\liml---equals zero.

The other special case in which the efficiency gain is zero is when $\delta=0$,
which by the Cauchy-Schwarz inequality, $\mu^{2}\leq \delta\Xi_{22}$, implies
$\mu=0$. The term $\delta$ measures the balance of the design matrices $Z$ and
$W$. If the diagonal elements of the projection matrices $(\Zp\Zp')_{ii}$ and
$(W(W'W)^{-1}W)_{ii}$, called the leverage of $i$, are constant across $i$, then
$\delta_{n}=0$, and $\delta_{n}$, and hence $\delta$, generally increases with
the variability of the leverages. Suppose, for instance, that each observation
$i$ belongs to one of $\Ni+1$ groups, numbered $0,\dotsc,\Ni$, and that there
are $n_{j}$ observations in group $j$. Let the $\Ni$-vector of instruments
$z^{*}_{i}$ correspond to a vector of group indicators, with group $0$ excluded,
and suppose that the only covariate is an intercept.\footnote{This setup arises
  when individuals are randomly assigned to groups. For example, if a defendant
  is randomly assigned to one of $\Ni+1$ judges who differ in their sentencing
  severity, then one can use judge indicators as instruments for the length of
  sentence of the defendant, as in \citet{AiDo15} or \citet{DoSo15}.} Then,
$(W(W'W)^{-1}W')_{ii}=1/n$, and $(ZZ')_{ii}=1/n_{j(i)}-1/n$, where $j(i)$
denotes the group that $i$ belongs to (see supplemental appendix for
derivation). It then follows from the definition of $\delta_{n}$ that
$\delta_{n}=0$ if and only if all groups have equal size, and that the magnitude
of $\delta_{n}$ increases with $\sum_{j=0}^{\Ni}1/n_{j}$, which can be thought
of as a measure of group size variability.

Recently, \citet{ccj12}, in the context of few strong instrument asymptotics,
proposed a modification of \liml\ that is more efficient than \liml\ when the
distribution of the reduced-form errors is not normal. The modification was to
use a more efficient estimator of the reduced-form coefficients $\Pin$ than
$\hat{\Pi}$. \citet{ccj12} use a two-step estimator that uses a kernel estimator
of the distribution of the reduced-form errors in the first step Under
\Cref{an:mi}, when the number of regressors in the reduced-form regression
increases with sample size however, this kernel estimator will not be
consistent, and so this estimator is unlikely to perform well in settings with
many instruments. In contrast, $\h{\beta}{emd}$ uses the same estimator
$\hat{\Pi}$ of $\Pin$ as \liml, but combines the information about $\beta$ in
$\hat{\Pi}$ in a more efficient way. On the other hand, $\h{\beta}{emd}$
requires $\alphai>0$ for the efficiency gain to be non-zero.

\section{Minimum distance estimation without rank
  restriction}\label{sec:missp--clust}

\Cref{an:pr} implies that the matrix $\Xi_{n}$ is
reduced rank. In particular, it implies that there are two sources of
information for estimating $\beta$,
\begin{align}
  \Xi_{11,n}&=\Xi_{12,n}\beta,\label{eq:id:1112}\qquad \text{and}\\
  \Xi_{12,n}&=\Xi_{22,n}\beta.\label{eq:id:1222}
\end{align}
The minimum distance objective function~\eqref{eq:md:objective2} weights both
sources of identification. In this section, I consider estimation without
imposing the rank restriction that
$\Xi_{11,n}/\Xi_{12,n}=\Xi_{12,n}/\Xi_{22,n}$. I show that a version of the
bias-corrected two-stage least squares estimator \citep{nagar59,dn01} is
equivalent to a minimum distance estimator that only uses~\eqref{eq:id:1222} to
estimate $\beta$, and derive standard errors that remain valid
when~\eqref{eq:id:1112} does not hold.

\subsection{Motivation for relaxing the rank restriction}

There are two important cases in which the ratios $\Xi_{12,n}/\Xi_{22,n}$ and
$\Xi_{11,n}/\Xi_{12,n}$, which correspond to estimands of the reverse two-stage
least squares and two-stage least squares estimators under standard asymptotics
\citep{kolesar13late}, are not necessarily equal to each other, but
$\Xi_{12,n}/\Xi_{22,n}$, is still of interest.

The first case arises when the effect of $x_{i}$ on $y_{i}$ is heterogeneous, as
in \citet{ia94}. Let $y_{i}(x)$ denote the potential outcome of individual $i$
when assigned $x_{i}=x$, and similarly let $x_{i}(z)$ denote the potential value
of the endogenous variable if the individual was assigned $z_{i}=z$. We observe
$y_{i}=y_{i}(x_{i})$ and $x_{i}=x_{i}(z_{i})$. For simplicity, suppose there are
no regressors $w_{i}$ beyond a constant. Suppose that
\begin{inparaenumi}
\item $z_{i}$ affects the outcome only through its effect on $x_{i}$:
  $\{y_{i}(x)\}_{x\in\mathcal{X}}$ is independent of $z_{i}$, where
  $\mathcal{X}$ denotes the support of $x_{i}$; and
\item Monotonicity holds: for any pair $(z_{1},z_{2})$,
  $\Prob(x_{i}(z_{1})\geq x_{i}(z_{2}))$ equals either zero or one.
\end{inparaenumi}
Then $\Xi_{12,n}/\Xi_{22,n}$ can be written as a particular weighted average of
average partial derivatives $\beta(z) =\E[\partial y_{i}(x_{i}(z))/\partial x]$
(see \citet{ai95} and \citet{agi00} for details). However, unless $\beta(z)$ is
constant, the rank restriction will not hold, and the ratio
$\Xi_{11,n}/\Xi_{12,n}$ may be outside of the convex hull of the average partial
derivatives \citep{kolesar13late}, which makes it hard to interpret.

The second case arises when instruments have a direct effect on the outcome. In
this case, the coefficient $\piss$ in the reduced-form regression of the outcome
on instruments is given by $\piss=\pifs\beta+\beta_{n}^{z}$, where
$\beta_{n}^{z}$ measures the strength of the direct effect. The structural
equation~\eqref{eq:structural-equation} no longer holds---instead we have
$y_{i}=x_{i}\beta+w_{i}'\beta_{n}^{w}+z_{i}'\beta_{n}^{z}+\epsilon_{i}$. Without
any restrictions on $\beta_{n}^{z}$, the parameter $\beta$ is no longer
identified. However, \citet{kcfgi14} show that if the direct effects are
orthogonal to the effects of the instruments on the endogenous variable in the
sense that $\pifs' \beta_{n}^{z}/n\to 0$ as $n\to\infty$, then $\beta$ can still
be consistently estimated. In particular, under this condition
$\Xi_{12,n}/\Xi_{22,n}=\beta+{\beta_{n}^{z}}'\pifs/\pifs'\pifs\to \beta$. In
contrast, $\Xi_{11,n}/\Xi_{12,n}\to \beta$ only if direct effects disappear
asymptotically so that ${\beta_{n}^{z}}'\beta_{n}^{z}/n\to 0$.

\subsection{Unrestricted minimum distance estimation}\label{sec:unrestr-minim-dist}
To relax the rank restriction on $\Xi_{n}$, define $\beta_{n}$ simply as the
ratio $\Xi_{12,n}/\Xi_{22,n}$, and consider the objective function
\begin{multline}\label{eq:md:ure}
  \mathcal{Q}^{\umd}_{n}(\beta,\Xi_{11},\Xi_{22};\hat W_{n})=\\
  \vech\left(T-(\Ni/n)S- \Bigl( \begin{smallmatrix}
      \Xi_{11}& \Xi_{22}\beta\\
      \Xi_{22}\beta& \Xi_{22}
  \end{smallmatrix}
  \Bigr)
\right)'\hat W_{n}
  \vech\left(T-(\Ni/n)S- \Bigl( \begin{smallmatrix}
      \Xi_{11}& \Xi_{22}\beta\\
      \Xi_{22}\beta& \Xi_{22}
  \end{smallmatrix}
  \Bigr)\right),
\end{multline}
where $\hat{W}_{n}\in\mathbb{R}^{3\times 3}$ is some weight matrix. If we
restrict $\Xi_{11,n}$ to equal to $\Xi_{22,n}\beta_{n}^{2}$, then minimizing
this objective function is equivalent to minimizing the original objective
function~\eqref{eq:md:objective2}. If $\Xi_{11,n}$ is unrestricted, the weight
matrix does not matter since then the model is exactly identified. The
unrestricted minimum distance estimators will be given by their sample
counterparts,
\begin{align*}
  \h{\Xi}{22,umd}&=T_{22}-(\Ni/n)S_{22},& \h{\Xi}{11,umd}&=T_{11}-(\Ni/n)S_{11},
\end{align*}
and
\begin{equation*}
  \h{\beta}{umd}=\frac{T_{12}-(\Ni/n)S_{12}}{T_{22}-(\Ni/n)S_{22}}.
\end{equation*}
The unrestricted minimum distance estimator for $\beta_{n}$ coincides with the
modified bias-corrected two-stage least squares estimator \citep{kcfgi14}, a
version of the bias-corrected two-stage least squares estimator. The version
proposed by \citet{dn01} multiplies $S_{12}$ and $S_{22}$ by
$\frac{\Ni-2}{n}\frac{\nus}{n-\Ni+2}$ instead of $\Ni/n$. The motivation for the
version in \citet{kcfgi14} was to modify the \citeauthor{dn01} estimator to make
it consistent when $\alphae >0$. However, it can also be viewed as a minimum
distance estimator that puts no restrictions on the reduced form. The next
proposition derives its large-sample properties.
\begin{proposition}\label{th:umd}
  Suppose that \Cref{an:mi}\ref{it:alphas}--\ref{it:z-lindeberg} and
  \Cref{an:rc} hold, $\Xi_{n}=\Xi+o(1)$ where $\Xi$ is a positive semi-definite
  matrix with $\Xi_{22}>0$, and that
  $(\piss-\pifs\beta_{n})'Z'\diag(H)/\sqrt{n\Ni}= \tilde{\mu}+o(1)$ for some
  $\tilde{\mu}$. Then
  \begin{equation*}
    \sqrt{n}\left(\h{\beta}{umd}-\beta_{n}\right)
    \indist\mathcal{N}(0,V_{\umd}),
  \end{equation*}
  where,
  \begin{equation*}
    \begin{split}
      V_{\umd}&=V_{\emd}+V_{\Delta}+\frac{\abs{\Xi}\Omega_{22}/\Xi_{22}
        +2\sqrt{\alphai}\tilde{\mu}\E[v_{2i}^{2}\epsilon_{i}]}{\Xi_{22}^{2}}\\
      V_{\Delta}&= \frac{\left((2\tau+\alphai\delta\kappa)\gamma (b'\Omega
          b)^{2}+ \sqrt{\alphai}\mu\E[\epsilon_{i}^{3}]
          +\alphai\delta\E[\epsilon_{i}^{3}(v_{2i}-\gamma
          \epsilon_{i})]\right)^2}{ (b'\Omega
        b)^{2}(2\tau+\alphai\delta\kappa)\Xi_{22}^{2} }
    \end{split}
  \end{equation*}
  where $\kappa$, $\gamma$ are defined as in~\eqref{eq:gamma-defn}
  and~\eqref{eq:ek}, with $\beta=\Xi_{12}/\Xi_{22}$, and
  $\epsilon_{i}=v_{2i}-v_{1i}\beta$.
\end{proposition}

The asymptotic variance $V_{\umd}$ corresponds to the (1,1) element of the
matrix ${G_{\umd}^{-1}}\Delta_{\umd} {G_{\umd}^{-1}}'$, where $G_{\umd}$ is the
derivative of the moment condition,
\begin{equation*}
  G_{\umd}=
  \begin{pmatrix}
    1 & 1 & 0\\
    \Xi_{22}& 0 & \beta\\
    0 & 0 & 1
  \end{pmatrix},\quad\text{so that}\quad {G_{\umd}^{-1}}'\left(
    \begin{smallmatrix}
      1\\0\\0
    \end{smallmatrix}
  \right)=\frac{1}{\Xi_{22}}(0,1,-\beta),
\end{equation*}
and, as shown in the proof,
$\Delta_{\umd}=L_{2}(\Delta_{1,\umd}+\Delta_{2}+
\Delta_{3,\umd}+\Delta_{3,\umd}')L_{2}'$ is the asymptotic variance of the
moment condition~\eqref{eq:md:moment-conditions:S-T}, with
\begin{align*}
  \Delta_{1,\umd}
  &=2N_{2}(\Xi\otimes \Omega+\Omega\otimes \Xi+\tau \Omega\otimes
              \Omega),
  & \Delta_{3,\umd}
  &=2\sqrt{\alphai}N_{2}\left(\Psi_{3}'\otimes
    \begin{pmatrix}
      \tilde{\mu}+\mu \beta\\\mu
    \end{pmatrix}\right),
\end{align*}
and $\Delta_{2}$ given in \Cref{th:moment-variance}. If $\Xi=\Xi_{22}aa'$, then
the expressions for $\Delta_{1,\umd}$ and $\Delta_{3,\umd}$ reduce to those for
$\Delta_{1}$ and $\Delta_{3}$ given in \Cref{th:moment-variance}.

The asymptotic variance consists of three components. The first term coincides
with the asymptotic variance of \emd\ given in~\Cref{eq:vemd}. The second
component, $V_{\Delta}$, represents the asymptotic efficiency loss relative to
$\h{\beta}{emd}$ when the rank restriction holds; it quantifies the price for
not using information contained in~\eqref{eq:id:1112} when the rank restriction
holds. Unlike the efficiency loss of \liml, the term is positive even when the
errors are normal, in which case it simplifies to
$2\tau (\Omega_{12}-\Omega_{22}\beta)^{2}/\Xi_{22}^{2}$, which is only zero if
there is no endogeneity (that is, $\E[x_{i}\epsilon_{i}]=0$). Finally, the last
component represents the increase in asymptotic variance due to the failure of
rank restriction; when~\ref{an:pr} holds, $\abs{\Xi}=0$ and $\tilde{\mu}=0$, and
this term drops out.

The asymptotic variance can be easily consistently estimated by
\begin{equation*}
  \hat{V}_{\umd}=\frac{1}{\hat{\Xi}_{22,\umd}^{2}}
  (0,1,-\h{\beta}{umd})'\hat{\Delta}_{\umd}
  (0,1,-\h{\beta}{umd}),
\end{equation*}
where $\h{\Delta}{umd}$ is a plug-in estimator based on
$\hat{\Xi}_{\umd}=T-\Ni/nS$, $\h{\beta}{umd}$, $\hat{\Omega}=S$, and estimators
of $\Psi_{3}$ and $\Psi_{4}$ given in \Cref{th:moment-variance}. Confidence
intervals based on $\h{\beta}{umd}$ and $\h{V}{\umd}$ will then be robust to
both many instruments, and failure of the proportionality
restriction~\eqref{eq:reduced-form}.

It is possible to reduce the asymptotic mean-squared error of the minimum
distance estimator by minimizing the minimum distance objective function subject
to the constraint that $\Xi_{n}$ be positive semi-definite (which has to be the
case since $\Xi_{n}$ is a matrix of second moments of $\pifs$ and $\piss$),
which is equivalent to the constraint $\Xi_{11,n}\geq \beta^{2}_{n}\Xi_{22,n}$.
If the weight matrix $\hat{W}_{n}$ is used, then the resulting estimator will
be a mixture between $\h{\beta}{umd}$, and the restricted minimum distance
estimator that minimizes~\eqref{eq:md:objective2} with respect to $\hat{W}_{n}$:
when $T-(\Ni/n)S$ is positive semi-definite, then the estimator equals
$\h{\beta}{umd}$; otherwise, the minimum distance objective is minimized at a
boundary, and the estimator equals the restricted minimum distance estimator.
When $\Xi_{n}$ is full rank, then the constraint won't bind in large samples,
and the estimator will be asymptotically equivalent to $\h{\beta}{umd}$.
However, when $\Xi_{n}$ is reduced-rank, the mixing will deliver a smaller
asymptotic mean-squared error. The disadvantage is that the estimator will be
asymptotically biased, which makes inference more complicated. See supplemental
appendix for details.

\section{Tests of overidentifying restrictions}\label{sec:specification-tests}

The proportionality restriction~\ref{an:pr} is testable. In this section, I
discuss a simple test based on the minimum distance objective function, and
compare it to some alternatives previously proposed in the literature.

In the invariant model, testing \Cref{an:pr} is equivalent to testing the null
that $\Xi_{n}$ is reduced-rank against the alternative that it is positive
definite. A simple way to implement the test is to compare the value of the
minimum distance objective function~\eqref{eq:md:ure} minimized subject to the
restriction that $\abs{\Xi_{n}}$ is reduced rank with its value when it is
minimized subject to $\abs{\Xi_{n}}$ being positive semi-definite. When
$\h{W}{re}$ is used as a weight matrix, the test statistic is given by (see
supplemental appendix for derivation)
\begin{equation*}
  \begin{split}
    \hat{J}_{\md}&=
    \min_{\Xi_{11}=\Xi_{22}\beta^{2}}\mathcal{Q}^{\umd}_{n}(\beta,\Xi_{11},\Xi_{22};\h{W}{re})-
    \min_{\Xi_{11}\geq\Xi_{22}\beta^{2}}
    \mathcal{Q}^{\umd}_{n}(\beta,\Xi_{11},\Xi_{22};\h{W}{re})\\
    &=\begin{cases}
      0 & \text{if $m_{\min}\leq \Ni/n$,} \\
      (m_{\min}-\Ni/n)^{2} & \text{otherwise.}
    \end{cases}
  \end{split}
\end{equation*}
The test statistic depends on the data through the minimum eigenvalue of
$S^{-1}T$. Because the weight matrix $\h{W}{re}$ is not optimal, the
large-sample distribution of $\h{J}{md}$ is not pivotal under the null: if
$\Ni\to\infty$ as $n\to\infty$, then in large samples, $n^{2}\h{J}{md}/\Ni$ will
be distributed as a mixture between a $\chi^{2}_{1}$ distribution scaled by
$\frac{2(1-\alphae)}{\alphanu}+\delta\kappa$ (with $\kappa$ defined in
\Cref{eq:ek}), and a degenerate distribution with a point mass at 0. One
solution would be to divide the test statistic by
$\frac{2(1-\alphae)}{\alphanu}+\delta\kappa$ and use a critical value based on
the 90\% quantile of a $\chi^{2}_{1}$ distribution, or, equivalently, reject
whenever
$(n/\sqrt{\Ni})(m_{\min}-\Ni/n)/\sqrt{\frac{2(1-\alphae)}{\alphanu}+\delta\kappa}$
is greater the 95\% quantile of a standard normal distribution. However, since
the asymptotic distribution changes when $\Ni$ is fixed, this won't yield valid
inference when $\Ni$ is fixed. Using arguments similar to \citet{ag11}, the next
proposition proposes a modification that ensures size control whether $\Ni$ is
fixed or grows with the sample size.
\begin{proposition}\label{th:overid}
  Suppose \Cref{an:pr,,an:mi,an:rc} hold. Suppose also that if $\Ni=K$ is
  fixed, then $\sup_{n\geq 1}\max_{i=1,\dotsc,n}(ZZ')_{ii}^{2}=o(1)$. Then:
  \begin{compactenumi}
  \item If $\Ni\to\infty$,
    $\frac{n}{\sqrt{\Ni}}(m_{\min}- \Ni/n) \indist \mathcal{N} (0,
    \frac{2(1-\alphae)}{\alphanu}+\delta\kappa)$. If $\Ni=K$ is fixed, then
    $nm_{\min}\indist \chi^{2}_{K-1}$.
  \item Let ${\delta}_{n}=\diag(H)'\diag(H)/\Ni$, let
    $\hat{\kappa}=(\h{b}{re}\otimes \h{b}{re})'\hat{\Psi}_{4}(\h{b}{re}\otimes
    \h{b}{re})/(\h{b}{re}'S\h{b}{re})^{2}-3$, and let $\Phi$ denote cdf of a
    standard normal distribution. The test that rejects whenever $nm_{\min}$ is
    greater than the
    \begin{equation*}
      1-\Phi\left(\textstyle\sqrt{\frac{(n-\Ni)}{\nus}+
          \frac{{\delta}_{n}\hat{\kappa}}{2}}\cdot\Phi^{-1}(\ns)\right)
    \end{equation*}
    quantile of the $\chi^{2}_{\Ni-1}$ distribution has asymptotic size equal to
    $\ns$. This holds whether $\Ni=K$ is fixed or $\Ni\to\infty$.
  \end{compactenumi}
\end{proposition}
When $\Ni\to\infty$, the test is asymptotically equivalent to the test proposed
in the previous paragraph. However, unlike that test, it also remains valid
under the few strong instrument asymptotics with $\Ni$ fixed. In this case, it
is asymptotically equivalent to the \citet{cd93} test, which is based on the
minimum distance objective function~\eqref{eq:md:go}, and rejects whenever
$nm_{\min}$ is greater than the $1-\ns$ quantile of $\chi^{2}_{\Ni}$
distribution. The test can therefore be interpreted as a Cragg-Donald test with
a modified critical value that ensures size control under few strong, as well as
many-instrument asymptotics.

It is interesting to compare this test to some other tests proposed in the
literature. In the context of few strong instrument asymptotics, the most
popular test is due to \citet{sargan58}. The test statistic can be written as
$\h{J}{s}=\frac{m_{\min}}{1-\Ni/n-\Ne/n+ m_{\min}}$, and the critical value is
given by $1-\ns$ quantile of $\chi^{2}_{\Ni}$. \citet{ag11} show that if
$\alphai>0$ and $\alphae=0$ and the errors are normal, the Sargan test is mildly
conservative. With $\alphai=0.1$ for example, the asymptotic size of the test
with nominal size $0.05$ is given by $0.04$. \citet{ag11} therefore propose an
adjustment to the critical value similar to the one proposed here to match the
asymptotic size with the nominal size. Unfortunately, this solution is not
robust to allowing the number of exogenous regressors to increase with the
sample size: if $\alphae>0$, the asymptotic size of the Sargan test converges to
one (see supplemental appendix for details). \citet{lo12} propose a different
modification of the Sargan test that controls size under conditions similar to
\Cref{th:overid}, provided that, in addition, $\alphae=0$ and $\Ni\to\infty$. In
contrast, the test proposed here will work irrespective of the number of
regressors or instruments; the researcher doesn't have to determine what type of
asymptotics are appropriate.

Another alternative to the test in \Cref{th:overid} would be to use the
efficient weight matrix instead of $\h{W}{re}$ in the minimum distance objective
function. Such a test would in general direct local asymptotic power to
different alternatives, and, without specifying which local violations of the
proportionality restriction are of interest, it is unclear which test should be
preferred. However, an attractive feature of the test in \Cref{th:overid} is its
easy implementation, which only requires modifying the critical value of the
Cragg-Donald test.

\section{Conclusion}\label{sec:conclusion}
In this paper, I outlined a minimum distance approach to inference in a linear
instrumental variables model with many instruments. I showed how
estimation and inference based on the minimum distance objective function solves
the incidental parameters problem that the large number of instruments create.
When the efficient weight matrix is used, I obtain a new estimator that is in
general more efficient than \liml. Moreover, depending on the weight matrix
used, and whether a proportionality restriction on the reduced-form coefficients
is imposed, the bias-corrected two-stage least squares estimator, the \liml\
estimator, and the random-effects estimator, which is shown to coincide with
\liml, are obtained as particular minimum distance estimators. Standard errors
can easily be constructed using the usual sandwich formula for asymptotic
variance of minimum distance estimators.

The invariance argument underlying the construction of the minimum distance
objective function relied on the assumption of homoscedasticity. It would be
interesting to explore in future work how this approach can be adapted to deal
with heteroscedasticity, and whether similar minimum distance construction can
be used in other models with an incidental parameters problem.

\clearpage

\begin{appendices}
  \numberwithin{lemma}{section} %
  \numberwithin{corollary}{section} %
  \section*{Appendix}
  \allowdisplaybreaks

  Appendix~\ref{sec:auxiliary-lemmata} states and proves some auxiliary Lemmata
  that are helpful for proving the main results. Proofs of lemmata and
  propositions stated in the text are given in Appendix~\ref{sec:proofs}.
  Throughout the appendix, I use the following simple identifies that follow
  from simple algebra. For any positive definite matrix
  $\Omega\in\mathbb{R}^{2\times 2}$, vectors $a=(\beta,1)'$ and $b=(1,-\beta)'$,
  $\beta\in\mathbb{R}$, and constants $c_{1},c_{2}$:
  \begin{subequations}\label{eq:equalities}
    \begin{align}
      \Qs(\beta,\Omega)+\Qt(\beta,\Omega)
      &=\trace(\Omega^{-1}T),\label{eq:equalities:qsqt}\\
      \abs*{\Omega}a\Omega^{-1}a
      &=b'\Omega b,\label{eq:equalities:aoa}\\
      \abs*{c_{1}T+c_{2} S }
      &=(c_{1}m_{\max}+c_{2})(c_{1}m_{\min}+c_{2})\abs*{S}.\label{eq:equalities:yy}
    \end{align}
  \end{subequations}

  \section{Auxiliary Lemmata}\label{sec:auxiliary-lemmata}
  \begin{lemma}\label{th:kronecker}
    \begin{inparaenumi}
    \item\label{it:kronecker-lnd} If for some invertible matrix
      $V\in\mathbb{R}^{d\times d}$, $N_{d}V=VN_{d}$, then
      $(L_{d}N_{d}VL_{d}')^{-1}=D_{d}'V_{d}^{-1}D_{d}$.
    \item For an invertible matrix $V\in\mathbb{R}^{d\times d}$, a vector
      $m\in\mathbb{R}^{d}$ and a constant $c$,
      \begin{equation*}
        \left(V\otimes V+c(mm')\otimes (mm')\right)^{-1}
        =V^{-1}\otimes V^{-1}-\frac{c(V^{-1}mm'V^{-1})\otimes
          (V^{-1}mm'V^{-1})}{1+c(m'V^{-1}m)^{2}},
      \end{equation*}

    \end{inparaenumi}
  \end{lemma}
  \begin{sproof}
    \begin{inparaenumi}
    \item It follows from Lemmata 3.5(i) and 3.6(ii) in \citet{mn80} that
      $L_{d}N_{d}D_{d}=I_{d(d+1)/2}$. Also, by Lemma 3.5(ii) in \citet{mn80},
      $D_{d}L_{d}N_{d}=N_{d}$. Thus,
      $(D_{d}'V^{-1}D_{d})(L_{d}N_{d}VL_{d}')=
      D_{d}'V^{-1}N_{d}VL_{d}'=D_{d}'V^{-1}VN_{d}L_{d}'
      =D_{d}'N_{d}L_{d}'=I_{d(d+1)/2}$.
    \item Follows from direct calculation.
    \end{inparaenumi}
  \end{sproof}
\begin{lemma}\label{th:clt}
  Consider the quadratic form $\Qf=(\Mm+\Em)'\Pm(\Mm+\Em)$, where
  $\Pm\in\mathbb{R}^{n\times n}$ is a symmetric matrix with non-random elements,
  $\Em,\Mm\in\mathbb{R}^{n\times G}$, $\Mm$ is non-random, and the rows
  $\ei{i}'$ of $\Em$ are i.i.d.\ with zero mean, variance $\Omega_{n}$, and
  finite fourth moments.

  Let $\Lambda_{n}=\Mm'\Pm\Pm \Mm$, $\delta_{n}=\diag(\Pm)'\diag(\Pm)$,
  $\overbar{m}_{n}=\Mm'\Pm\diag(\Pm)$, $p_{ij}=(\Pm)_{ij}$, and let $e_{in}$
  denote an $n$-vector of zeros with $1$ in the $i$th position. Then
  \begin{compactenumi}
  \item\label{it:variance} The variance of $\Qf$ is given by
    \begin{multline*}
      \var(\mkvec(\Qf)) = 2N_{G}\left(\Omega_{n}\otimes \Lambda_{n}+\Lambda_{n}\otimes
        \Omega_{n}
        +\trace(\Pm^{2})\Omega_{n}\otimes \Omega_{n}  \right)\\
      +\delta_{n}\left(\E[\ei{i}\ei{i}'\otimes
        \ei{i}\ei{i}']-\mkvec(\Omega_{n})\mkvec(\Omega_{n})'
        -2N_{G}\Omega_{n}\otimes \Omega_{n}
      \right)\\
      +2N_{G}(\Psi_{3}'\otimes\overbar{m}_{n})
      +2(\Psi_{3}'\otimes\overbar{m}_{n})'N_{G},
    \end{multline*}
    where $\Psi_{3}=E[(\ei{i}\ei{i}')\otimes \ei{i}]$, and the last two lines
    are equal to zero if the distribution of $\ei{i}$ is normal.
  \item\label{it:centr-limit-theor} Suppose that
    \begin{inparaenumi}
    \item\label{it:convergences} $\var(\mkvec(\Qf))$ converges, and $\delta_{n}$
      and $\Lambda_{n}$ are bounded;
    \item\label{it:four-plus-delta-moments}
      $\sup_{n}\E[\norm{\ei{i}}^{8}]<\infty$;
    \item\label{it:P-conditions} $\sum_{i=1}^{n}\abs{p_{ii}}^{4}=o(1)$,
      $\sum_{i=1}^{n}\big(\sum_{j=1}^{n} p_{ij}^{2} \big)^{2}=o(1)$, and
      $\sum_{i<j<k<\ell}^{} p_{ik}p_{i \ell } p_{jk} p_{j \ell }=o(1)$; and
    \item\label{it:M-condition} $\sum_{i=1}^{n}\norm{e_{in}'\Pm \Mm}^{4}=o(1)$.
    \end{inparaenumi}
    Then:
    \begin{equation*}
      \mkvec(\Qf-\Mm'\Pm\Mm-\trace(\Pm)\Omega_{n})\indist
      \mathcal{N}(0,\lim_{n\to\infty}(\var(\mkvec(\Qf)))).
    \end{equation*}
  \end{compactenumi}
\end{lemma}
\begin{sproof}
  Proof of Part~\ref{it:variance} follows from a tedious, but straightforward
  calculation. Proof of Part~\ref{it:centr-limit-theor} is a generalization of
  the central limit theorems in \citet{cshnw12} and \citet{hhn08}, and is proved
  using similar arguments. Full proof is given in the supplemental appendix.
\end{sproof}
\begin{lemma}\label{th:lemma-p4}
  Let $\Pm=(A_{n}+\nu_{n}B_{n})/\sqrt{m_{n}}$, where $m_{n}\to\infty$ as
  $n\to\infty$, $\nu_{n}=O(1)$, $A_{n},B_{n}\in\mathbb{R}^{n\times n}$ are
  projection matrices such that $A_{n}B_{n}=0$, and for $j>1$,
  $\trace(A_{n}/m_{n}^{j})=o(1)$ and $\trace(\nu_{n}B_{n}/m_{n}^{j})=o(1)$. Then
  condition~\ref{it:P-conditions} of \Cref{th:clt} holds.
\end{lemma}
\begin{sproof}
  Denote the $(i,j)$ elements of $A_{n}$ and $B_{n}$ by $a_{ij}$ and $b_{ij}$.
  The first condition follows from the bound, for $j> 2$,
  $\sum_{i}p_{ii}^{j}\leq
  2^{j-1}(\sum_{i}a_{ii}^{j}+\nu_{n}^{j}\sum_{i}b_{ii}^{j})/m_{n}^{j/2}\leq
  2^{j-1}(\sum_{i}a_{ii}+\nu_{n}^{j}\sum_{i}b_{ii})/m_{n}^{j/2}=o(1)$ The second
  condition follows from
  $\sum_{i}\big(\sum_{j}p_{ij}^{2}\big)^{2}\leq
  \sum_{i}\big(2\sum_{j}a_{ij}^{2}+2\nu_{n}^{2}\sum_{j}b_{ij}^{2}\big)^{2}/m_{n}^{2}
  =\sum_{i}\big(2a_{ii}+2\nu_{n}^{2}b_{ii}\big)^{2}/m_{n}^{2}=o(1)$.

  It therefore remains to show that
  $\sum_{i<j<k<\ell}p_{ik}p_{i\ell}p_{jk}p_{j\ell}=o(1)$. This can be shown
  using arguments similar to those in the proof of Lemma B.2 in \citet{cshnw12}.
  Let $D$ denote a diagonal matrix with elements $D_{ii}=(\Pm)_{ii}$, let
  $S_{n}=\sum_{i<j<k<\ell}(p_{ik}p_{i\ell}p_{jk}p_{j\ell}
  +p_{ij}p_{i\ell}p_{jk}p_{k\ell} +p_{ij}p_{ik}p_{j\ell}p_{k\ell} )$, and let
  $\norm{\cdot}_{F}$ denote the Frobenius norm. Note that
  \begin{equation}\label{eq:trace-pd}
    \norm{(\Pm-D)^{2}}_{F}\leq\norm{\Pm^{2}}_{F}+\norm{D^{2}}_{F}
    +2\norm{D\Pm}_{F}
    =o(1),
  \end{equation}
  where the last equality follows from
  $\norm{D^{2}}^{2}_{F}=\sum_{i}p_{ii}^{4}=o(1)$,
  $\norm{\Pm^{2}}_{F}^{2}=(\trace{A_{n}}+\nu_{n}^{4}\trace{B_{n}})/m_{n}^{2}=o(1)$,
  and
  $\norm{D\Pm}_{F}^{2}=\sum_{i}p_{ii}^{2}\sum_{j}p_{ij}^{2} \leq
  m_{n}^{-2}\sum_{i}(a_{ii}+\nu_{n} b_{ii})(a_{ii}+\nu_{n}^{2}b_{ii}) =o(1)$. On
  the other hand, expanding the left-hand side in~\eqref{eq:trace-pd} yields
  \begin{equation*}
    \norm{(\Pm-D)^{2}}^{2}_{F}=2\sum_{i<j}p_{ij}^{4}
    +4\sum_{i<j<\ell}\left(p_{ij}^{2}p_{i\ell}^{2}
      +  p_{ij}^{2}p_{j\ell}^{2}+p_{j\ell}^{2}p_{i\ell}^{2}\right)+ 8S_{n}.
  \end{equation*}
  Since the first four terms are bounded by
  $\sum_{i}(\sum_{j}p_{ij}^{2})^{2}=o(1)$, it follows that $S_{n}=o(1)$.
  Define
  \begin{align*}
    \Delta_{2}& =\sum_{i<j<k}\left(p_{ij}p_{ik}\epsilon_{j}\epsilon_{k}
                +p_{ij}p_{jk}\epsilon_{i}\epsilon_{k}
                \right),\\
    \Delta_{3}&=\sum_{i<j<k}\left(p_{ik}p_{jk}\epsilon_{i}\epsilon_{j} \right),
  \end{align*}
  and let $\Delta_{1}=\Delta_{2}+\Delta_{3}$. Then
  \begin{align*}
    \E[\Delta_{3}^{2}]&=\sum_{i<j<k,\ell}p_{ik}p_{jk}p_{i\ell}p_{j\ell} =
                        \sum_{i<j<k}p_{ik}^{2}p_{jk}^{2}+
                        2\sum_{i<j<k<\ell}p_{ik}p_{jk}p_{i\ell}p_{j\ell}
                        =2\sum_{i<j<k<\ell}p_{ik}p_{jk}p_{i\ell}p_{j\ell}+o(1),\\
    \E[\Delta_{1}^{2}]&=\trace((\Pm-D)^{4})-2\sum_{i<j}p_{ij}^{4}=o(1),\\
    \E[\Delta_{2}^{2}]&=\sum_{i<j<k}(p_{ij}^{2}p_{ik}^{2}+p_{ij}^{2}p_{jk}^{2})
                        +2S_{n}=o(1).
  \end{align*}
  Thus, by the Cauchy-Schwarz inequality,
  \begin{equation*}
    \E[\Delta_{3}^{2}]\leq 2\E[\Delta_{1}^{2}]+2\E[\Delta_{2}^{2}]=o(1),
  \end{equation*}
  which proves the result.
\end{sproof}
  \begin{corollary}\label{th:st:adist:valid}
    Consider the model~\eqref{eq:reduced-form}--\eqref{eq:homo}, and suppose
    \Cref{an:pr,,an:n,an:mi} hold. Then:
    \begin{align*}
      \sqrt{n}\mkvec\left(S-\Omega\right)
      &\indist\mathcal{N}_{4}\left(0,\frac{1}{1-\alphai-\alphae}
        2N_{2}(\Omega\otimes
        \Omega) \right)\\
      \sqrt{n}\mkvec\left(T-\alphai
      \Omega-\frac{\lambdan}{a'\Omega^{-1}a}aa'\right)
      &\indist\mathcal{N}_{4}\left(0,
        2N_{2}(\alphai \Omega\otimes \Omega+\Omega\otimes M
        +M\otimes\Omega)\right),
    \end{align*}
    where $M=\frac{\lambda}{a'\Omega^{-1}a}aa'$.
  \end{corollary}
  \begin{sproof}
    The result follows from~\Cref{th:clt,th:lemma-p4}, with
    $P_{n}=(I-ZZ'-W(W'W)^{-1}W)/\sqrt{n}$, and $P_{n}=(ZZ')/\sqrt{n}$.
  \end{sproof}
\begin{corollary}\label{th:moment-distro}
  Consider the model~\eqref{eq:reduced-form}--\eqref{eq:homo}, and suppose
  \Cref{an:mi}\ref{it:alphas}--\ref{it:z-lindeberg}, and \Cref{an:rc} hold,
  $\Xi_{n}=\Xi+o(1)$ where $\Xi$ is a positive semi-definite matrix with
  $\Xi_{22}>0$, and that
  $(\piss-\pifs\beta_{n})'Z'\diag(H)/\sqrt{n\Ni}= \tilde{\mu}+o(1)$ for some
  $\tilde{\mu}$. Let $\overbar{m}=(\tilde{\mu}+\mu(\Xi_{12}/\Xi_{22}),\mu)'$.
  Then
  \begin{equation*}
    \sqrt{n}\mkvec(T-(\Ni/n)S-\Xi_{n})\indist\mathcal{N}(0,\Delta),
  \end{equation*}
  where
  \begin{multline*}
    \Delta= 2N_{2}\left(\Omega\otimes \Xi+\Xi\otimes \Omega
      +(\alphai(1-\alphae)/(\alphanu)-\alphai\delta)\Omega\otimes \Omega  \right)\\
    +\alphai\delta\left(\E[v_{i}v_{i}'\otimes
      v_{i}v_{i}']-\mkvec(\Omega)\mkvec(\Omega)'\right)
    +\alphai^{1/2}(2N_{2}(\Psi_{3}'\otimes\overbar{m}) +
    (\Psi_{3}'\otimes\overbar{m})'N_{2}),
    \end{multline*}
    and $\Psi_{3}=E[(v_{i}v_{i}')\otimes v_{i}]$.
\end{corollary}
\begin{sproof}
    The result follows from~\Cref{th:clt,th:lemma-p4}, with
    $P_{n}=H/\sqrt{n}$.
\end{sproof}

\section{Proofs}\label{sec:proofs}
\begin{sproof}[\Cref{th:invariant-lik}]
  To ensure that the densities of $T$ and $\hat{\Pi}$ are expressed with respect
  to compatible dominating measures, I will express the density of $T$ with
  respect to the measure
  \begin{equation*}
    \mu_{T}(\dd t)
    =\frac{n^{\Ni}\pi^{\Ni-1/2}}{\Gamma(\Ni/2)\Gamma((\Ni-1)/2)}
    \abs{t}^{(\Ni-3)/2}\lambda(\dd t),
  \end{equation*}
  where $\lambda$ is the Lebesgue measure on the sample space of $T$, and
  $\Gamma$ denotes the gamma function. $\mu_{T}$ is the measure induced by the
  Lebesgue measure $\mu$ on the sample space of $\hat{\Pi}$ in the sense that
  for any measurable set $B$, $\mu_{T}(B)=\mu(\delta^{-1}(B))$, where
  $\delta(\hat{\Pi})=\hat{\Pi}'\hat{\Pi}/n$ is the function defining $T$
  \citep[Example 5.1]{eaton89}. The statistic $T$ has the same distribution as
  the statistic $W_{N}$ in \citet[Section 4]{moreira09as}, with the parameters
  $\lambda_{N}$ and $\Sigma$ in that paper corresponding to
  $\lambdan/(a'\Omega^{-1}a)$ and $\Omega$. Hence, by Theorem 4.1 in
  \citet{moreira09as}, the density of $T$ with respect to $\mu_{T}$ is given by
  \begin{equation}\label{eq:fT}
    f_{T}(T\mid \beta,\lambdan,\Omega)
    =\mathcal{K}_{1}
    e^{-\frac{n}{2}(\lambdan+\trace(\Omega^{-1}T))} \abs{\Omega}^{-\Ni/2}
    (n\lambdan^{1/2}\Qt(\beta,\Omega)^{1/2})^{-\frac{\Ni-2}{2}}I_{(\Ni-2)/2}(
    n\lambdan^{1/2}\Qt(\beta,\Omega)^{1/2}),
  \end{equation}
  where $\mathcal{K}_{1}=\Gamma(\Ni/2)\pi^{-\Ni}2^{-\Ni/2-1}$ and $I_\nu(\cdot)$
  is modified Bessel function of the first kind of order $\nu$. $I_\nu(\cdot)$
  has the integral representation \citep[Equation 9.6.18, p.~376]{as64}
  \begin{align*}
    I_{\nu}(t)
    &=\frac{(t/2)^{\nu}}{\pi^{1/2}{\Gamma}(\nu+1/2)}G_{2\nu+2}(t),
    &\text{where}\qquad G_{k}(t) =
      \int_{[-1,1]}e^{ts }(1-s^{2})^{(k-3)/2}\,\dd s.
  \end{align*}
  The density~\eqref{eq:fT} can therefore be written as:
  \begin{equation*}
    f_{T}(T\mid \beta,\lambdan,\Omega)
    =
    \frac{2^{-\Ni}\Gamma(\Ni/2)}{\pi^{\Ni+1/2}{\Gamma}((\Ni-1)/2)}\cdot
    e^{-\frac{n}{2}(\lambdan+\trace(\Omega^{-1}T))} \abs{\Omega}^{-\Ni/2}
    G_{\Ni}(
    n\lambdan^{1/2}\Qt(\beta,\Omega)^{1/2}).
  \end{equation*}
  Combining this expression with the density for $S$ (with respect to Lebesgue
  measure), which is given by
  \begin{equation*}
    f_{S}(S;\Omega)=
    C_{\nus}\cdot \abs{S}^{(\nus-3)/2}
    \abs{\Omega}^{-\nus/2}e^{-\frac{\nus}{2}\trace(\Omega^{-1}S)},
  \end{equation*}
  where
  \begin{equation}\label{eq:Cnu}
    C_{\nu}^{-1}=(2/\nu)^{\nu}\pi ^{1/2}\Gamma (\nu/2)\Gamma ((\nu-1)/2)
  \end{equation}
  yields the invariant likelihood
  \begin{equation}\label{eq:invariant-lik}
    \begin{split}
      \mathcal{L}_{\text{\an{inv}},n}(\beta,\lambdan,\Omega;S,T) &=
      \frac{2^{-\Ni}\Gamma(\Ni/2)}{\pi^{\Ni+1/2}{\Gamma}((\Ni-1)/2)}\cdot
      e^{-\frac{n}{2}(\lambdan+\trace(\Omega^{-1}T))} \abs{\Omega}^{-\Ni/2}
      G_{\Ni}( n\lambdan^{1/2}\Qt(\beta,\Omega)^{1/2})\cdot f_{S}(S;\Omega)
      \\
      &\propto \exp\left[-\frac{1}{2}\left( (n-\Ne ) \log\abs*{\Omega}
          +\trace(\Omega^{-1}\tilde{S} )+n\lambdan-2\log G_{\Ni
          }(n\sqrt{\lambdan \Qt(\beta,\Omega)})\right)\right],
    \end{split}
  \end{equation}
  where $\tilde{S}=(\nus)S+nT$. The derivative with respect to $\Omega$ is given
  by:
  \begin{equation}\label{eq:il:score-omega}
    \frac{\partial
      \log
      \mathcal{L}_{\text{\an{inv}},n}}{\partial \Omega}=
    \frac{1}{2}\Omega^{-1}\left[
      \tilde{S}- (n-\Ne )
      \Omega-\frac{G_{\Ni}'(
        n\sqrt{\lambdan \Qt(\beta,\Omega)})}{G_{\Ni}(
        n\sqrt{\lambdan \Qt(\beta,\Omega)})}
      \frac{n\lambdan^{1/2}}{
        \Qt(\beta,\Omega)^{1/2}}
      \left(T-\frac{\Qs(\beta,\Omega)}{b'\Omega
          b}\Omega bb'\Omega\right)\right]\Omega^{-1},
  \end{equation}
  where the derivative $\partial \Qt(\beta,\Omega)/\partial \Omega$ is computed
  using the identity~\eqref{eq:equalities:qsqt}. Note that $b'\Omega b>0$ for
  $\Omega$ positive definite, $\Qt(\beta,\Omega)>0$ with probability one for
  $\Omega$ positive definite, and ${G}_{\Ni}(t)>0$ for $t>0$
  \citep[][p.~374]{as64}, so that the denominators in~\eqref{eq:il:score-omega}
  are non-zero at any point in the parameter space for $(\beta,\Omega,\lambdan)$
  with probability one.

  Fix $\lambdan$. Denote the \an{ml} estimates of $\beta$ and $\Omega$ given
  $\lambdan$ by $(\hat\beta_{\lambdan},\hat\Omega_{\lambdan})$. Since $G(\cdot)$
  is a monotone increasing function, it follows from~\eqref{eq:invariant-lik}
  that:
  \begin{equation}\label{eq:betalambda}
    \hat\beta_{\lambdan}=\argmax_{\beta}\Qt(\beta,\hat\Omega_{\lambdan})=
    \argmin_{\beta}\Qs(\beta,\hat\Omega_{\lambdan}).
  \end{equation}
  Secondly, the derivative~\eqref{eq:il:score-omega} evaluated at
  $(\hat\beta_{\lambdan},\hat\Omega_{\lambdan})$ has to be equal to zero.
  Pre-multiplying and post-multiplying \Cref{eq:il:score-omega} by
  $\hat b_{\lambdan}'\hat\Omega_{\lambdan}$ and
  $\hat\Omega_{\lambdan} \hat{b}_{\lambdan}$ therefore yields
  \begin{equation}\label{eq:i:propyy}
    (n-\Ne )\hat b_{\lambdan}'\hat\Omega_{\lambdan}\hat b_{\lambdan}=
    \hat b_{\lambdan}    '\tilde{S} \hat b_{\lambdan}.
  \end{equation}
  Therefore,
  \begin{equation*}
    \hat{\beta}_{\lambdan}=\argmin_{\beta}
    \Qs(\beta,\hat{\Omega}_{\lambdan})=
    \argmin_{\beta}
    \Qs(\beta,\tilde{S})=\argmin \Qs(\beta,S)=\h{\beta}{liml},
  \end{equation*}
  where the first equality follows by~\eqref{eq:betalambda}, the second
  by~\eqref{eq:i:propyy}, the third by
  $\Qs(\beta,\tilde{S})^{-1}=(\nus)\Qs(\beta,S)^{-1}+n$, and the last equality
  follows from definition of $\h{\beta}{liml}$. By similar arguments,
  \Cref{eq:betalambda,eq:i:propyy} must also hold when the likelihood is
  maximized over $\lambdan$ as well, so that $\h{\beta}{inv}=\h{\beta}{liml}$.

  It remains to show~\eqref{eq:invariant-lik-integrated}. This result follows
  from the fact that $F_{\omegan}$ corresponds to the invariant prior
  distribution induced by the Haar probability measure $\nu_{H}$ on
  $\mathcal{O}(\Ni)$ (which is unique since $\mathcal{O}(\Ni)$ is compact) via
  the group action $\omegan\mapsto g\omegan$, $g\in\mathcal{O}(\Ni)$, in the
  sense that for any measurable set $B$, $F_{\omegan}(B)=\nu_{H}(g^{-1}B)$, and
  arguments in \citet[pp.~87--88]{eaton89}. For convenience, I give a direct
  argument. Since
  \begin{equation*}
    \mkvec(\hat\Pi)\sim \mathcal{N}_{2\Ni}\left((a'\Omega^{-1}a/n)^{-1/2} a\otimes
      \etan ,\Omega\otimes
      I_{\Ni}\right),
  \end{equation*}
  it follows that the limited information likelihood is given by
  \begin{equation*}
    \mathcal{L}_{\li,n}(\beta,\omegan,\lambdan,\Omega)=
    (2\pi)^{-\Ni}   \abs{\Omega}^{-{\frac {\Ni}{2}}}
    e^{-{\frac {n}{2}}
      \left(\trace(T\Omega^{-1})+\lambdan\right)}
    e^{
      n\sqrt{\lambdan \Qt(\beta,\Omega)}
      \omegan'A(\beta,\Omega,\hat{\Pi})}f_{S}(S;\Omega) ,
  \end{equation*}
  where
  $A(\beta,\Omega,\hat{\Pi})=\frac{\hat{\Pi}\Omega^{-1}a}{ (n a'\Omega^{-1}a
    \Qt(\beta,\Omega))^{1/2}}$. To integrate the likelihood, we use the result
  that for all $t\in\mathbb{R}$, $\alpha\in\mathbb{S}^{\Ni-1}$, and $\Ni\geq 2$
  \citep[see][pp.~88--89]{stroock99}
  \begin{equation*}
    \int_{\mathbb{S}^{\Ni-1}} e^{t\alpha'\omega}\,\dd F_{\omegan}(\omega)
    =\frac{\Gamma(\Ni/2)}{\pi^{1/2}\Gamma((\Ni-1)/2)}G_{\Ni}(t),
  \end{equation*}
  Applying this result with $t=n\sqrt{\lambdan \Qt(\beta,\Omega)}$ and
  $\alpha=A(\beta,\Omega,\hat{\Pi})$ gives
  \begin{equation*}
    \int
    \mathcal{L}_{\li,n}(\beta,\omega,\lambdan,\Omega)\,\dd F_{\omegan}(\omega)\\
    = \frac{2^{-\Ni}\Gamma(\Ni/2)}{\pi^{\Ni+1/2}\Gamma((\Ni-1)/2)}
    \abs{\Omega}^{-{\frac {\Ni}{2}}} e^{-{\frac {n}{2}}
      \left(\trace(T\Omega^{-1})+\lambdan\right)} G_{\Ni}(n\lambdan^{1/2}
    \Qt(\beta,\Omega)^{1/2})\cdot f_{S}(S;\Omega),
  \end{equation*}
  which in view of~\eqref{eq:invariant-lik} completes the proof.
\end{sproof}

\begin{sproof}[\Cref{th:re-theorem}]
  Let $\nu=\nus$, and to prevent clutter, I use the notation
  $(\hat{\beta},\hat{\lambda},\hat{\Omega})$ rather than
  $(\h{\beta}{re},\h{\lambda}{re},\h{\Omega}{re})$. Consider first maximizing
  the likelihood with respect to $\Omega$, holding $\beta$ and $\lambda$ fixed.
  Let $\hat{\Omega}_{\beta,\lambda}$ denote the resulting estimator. The
  derivative of the log-likelihood with respect to $\Omega$ is given by
  \begin{equation*}
    \frac{\partial \log \mathcal{L}_{\re, n} (\beta,\lambda,\Omega)
    }{\partial\Omega}    = \frac{1}{2}\left[
      \Omega^{ - 1}\tilde{S} \Omega^{ - 1} - (n - \Ne )\Omega^{ - 1} -
      d(\lambda)\left( \Omega^{ - 1}T\Omega^{ - 1} -
        \frac{\Qs(\beta,\Omega)}{
          b'\Omega b}bb'\right)\right],
  \end{equation*}
  where $\tilde{S}=nT+\nu S$ and $d(\lambda)=\frac{n\lambda}{\Ni / n + \lambda}$
  and the derivative $\partial \Qt(\beta,\Omega)/\partial \Omega$ is computed
  using the identity~\eqref{eq:equalities:qsqt}. Since the derivative equals
  zero at $\hat{\Omega}_{\beta,\lambda}$, this implies
  \begin{equation}\label{eq:omega-derivative}
    \hat{\Omega}_{\beta,\lambda}^{ - 1}\tilde{S} =
    (n - \Ne )I_{2}+
    d(\lambda)\left(     \hat{\Omega}_{\beta,\lambda}^{ - 1}
      T -
      \frac{\Qs(\beta,\hat{\Omega}_{\beta,\lambda})}{
        b'\hat{\Omega}_{\beta,\lambda} b}bb'\hat{\Omega}_{\beta,\lambda}\right).
  \end{equation}
  Taking a trace on both sides of the equation and using the
  identity~\eqref{eq:equalities:qsqt} then yields
  \begin{equation}\label{eq:trace-omega}
    \trace(\hat{\Omega}_{\beta,\lambda}^{-1}\tilde{S})
    -d(\lambda)\Qt(\beta,\hat{\Omega}_{\beta,\lambda})=2(n-\Ne).
  \end{equation}
  Pre- and post-multiplying~\Cref{eq:omega-derivative} by
  $b'\hat{\Omega}_{\beta,\lambda}$ and $b$; and by
  $\hat{\Omega}_{\beta,\lambda}$ and $b$ yields
  \begin{align*}
    b'\hat{\Omega}_{\beta,\lambda}b
    &= b'\tilde{S}b/(n-\Ne),\\
    (n - \Ne )\hat{\Omega}_{\beta,\lambda}b
    &=\frac{1}{1-d(\lambda) \Qs(\beta,\tilde{S})}
      (\tilde{S}-d(\lambda) T)b.
  \end{align*}
  Plugging these expressions back into~\Cref{eq:omega-derivative} and
  pre-multiplying the resulting expression by $\hat{\Omega}_{\beta,\lambda}$
  yields
  \begin{equation}\label{eq:hO-lb}
    (n - \Ne )  \hat{\Omega}_{\beta,\lambda}=
    \tilde{S}-d(\lambda)T +
    \frac{1}{
      b'(\tilde{S}-d(\lambda)T) b}
    \frac{d(\lambda)\Qs(\beta,\tilde{S})}{1-d(\lambda) \Qs(\beta,\tilde{S})}
    (\tilde{S}-d(\lambda) T)bb'
    (\tilde{S}-d(\lambda) T).
  \end{equation}
  Taking a determinant on both sides of the equation, using the matrix
  determinant lemma $\abs{A+cVV'}=\abs{A}(1+cV'A^{-1}V)$ with
  $A=\tilde{S}-d(\lambda) T$ and $V=(\tilde{S}-d(\lambda) T)b$, and the
  identity~\eqref{eq:equalities:yy} yields
  \begin{multline*}
    (n - \Ne )^{2}\abs{\hat{\Omega}_{\beta,\lambda}}=
    \frac{\abs{\tilde{S}-d(\lambda) T}}{ 1-d(\lambda) \Qs(\beta,\tilde{S})} =
    \frac{\abs{\nu S + \Ni/(\Ni/n+\lambda)\cdot T}}{
      \Ni/n+\lambda\nu/(\nu+n\Qs(\beta,S))}(\Ni/n+\lambda)
    \\
    =\frac{\abs{S}}{\Ni/n+\lambda} \frac{ (\Ni m_{\max}+\nu(\Ni/n+\lambda))(\Ni
      m_{\min}+\nu(\Ni/n+\lambda)) }{ \Ni/n+\lambda\nu/(\nu+n\Qs(\beta,S))}
  \end{multline*}
  Plugging this expression and the expression~\eqref{eq:trace-omega} back into
  the likelihood then yields that the log-likelihood with $\Omega$ concentrated
  out is given by
  \begin{equation}\label{eq:L-O-conc}
    \log\mathcal{L}_{\re,n}(\lambda,{\beta},\hat{\Omega}_{\beta,\lambda})
    \propto -\frac{1}{2}\left[
      \Ni\log \left(\frac{\Ni}{n} +
        \lambda\right)+ (n-\Ne)
      \log\left(
        \frac{
          (\Ni(m_{\max}+\frac{\nu}{n})
          +\nu\lambda)
          (\Ni(m_{\min}+\frac{\nu}{n})
          +\nu\lambda) }{(\Ni/n+\lambda)
          (\Ni/n+{\lambda\nu}/{(\nu+n\Qs(\beta,S))})}
      \right)\right].
  \end{equation}
  Since this expression depends on $\beta$ only through $\Qs(\beta,S)$, and is
  decreasing in $\Qs(\beta,S)$ for any $\lambda>0$, it follows that the maximum
  likelihood estimate of $\beta$ with $\lambda$ fixed at any positive value is
  given by $\hat{\beta}_{\lambda}=\argmin_{\beta} \Qs(\beta,S)=\h{\beta}{liml}$.
  If $\lambda=0$, then the expression doesn't depend on $\beta$, and we can in
  particular set $\hat{\beta}_{\lambda=0}=\h{\beta}{liml}$, so that
  $\hat{\beta}=\h{\beta}{liml}$. The log-likelihood with $\Omega$ and $\beta$
  both concentrated out is thus given by
  \begin{equation*}
    \log\mathcal{L}_{\re,n}(\lambda,\hat{\beta}_{\lambda},
    \hat{\Omega}_{\hat{\beta}_{\lambda},\lambda})
    \propto -\frac{1}{2}\left[
      \Ni\log \left(\Ni/n +
        \lambda\right)+ (n-\Ne)
      \log\left(
        \frac{ \Ni m_{\max} }{\Ni/n+\lambda
        }+    \nu
      \right)\right].
  \end{equation*}
  The derivative
  equals zero at $\lambda=m_{\max}-\Ni/n$, and is negative for
  $\lambda>m_{\max}-\Ni/n$, which implies that
  $\hat{\lambda}=\max\{m_{\max}-\Ni/n, 0\}$. Plugging in the expressions for
  $\hat{\lambda}$ and $\hat{\beta}$ into~\eqref{eq:hO-lb} then yields
  \begin{equation*}
    (n - \Ne )  \hat{\Omega}=
    \tilde{S}-d(\hat{\lambda})\left(T -
      \frac{1}{\hat{b}'T \hat{b}} T\hat{b}\hat{b}'T\right)
    =\tilde{S}-d(\hat{\lambda})\frac{\hat{a}\hat{a}'\abs{T}}{
      \hat{b}'T\hat{b}
    }=\tilde{S}-d(\hat{\lambda})m_{\max}
    \frac{\hat{a}\hat{a}'}{\hat{a}'S^{-1}\hat{a}},
  \end{equation*}
  where $\hat{b}=(1,-\hat{\beta})'$, the first equality uses the identity
  $T\hat{b}=m_{\min}S\hat{b}$, the second equality uses the identity
  $b'Mb\cdot M=\hat{a}\hat{a}'\abs{B}+Mbb'M$ that holds for any matrix $M$, and
  the last equality uses $\hat{b}'T\hat{b}=m_{\min}\hat{b}'S\hat{b}$
  and~\eqref{eq:equalities:aoa}.

  Next I derive the inverse Hessian. Let $e_{2}=(0,1)'$. The score equations
  based on the \re\ likelihood~\eqref{eq:re:lik} are given by:
  \begin{align}
    \mathcal{S}_{\beta}(\beta, \lambda, \Omega)
    &= d(\lambda)
      \frac{e_{2}'\left(T-\Qs(\beta,\Omega)\Omega\right)b}{b'\Omega b},\\
    \mathcal{S}_{\lambda}(\beta, \lambda, \Omega)
    &= - \frac{1}{2}\frac{\Ni
      }{\Ni / n + \lambda} \left(1 - \frac{\Qt(\beta, \Omega) }{\Ni / n +
      \lambda}\right),\\
    \mathcal{S}_{\Omega}(\beta, \lambda, \Omega)
    &= \frac{1}{2}D_{2}'\mkvec\left[
      \Omega^{ - 1}\tilde{S} \Omega^{ - 1} - (n - \Ne )\Omega^{ - 1} -
      d(\lambda)\left( \Omega^{ - 1}T\Omega^{ - 1} -
      \frac{\Qs(\beta,\Omega)}{b'\Omega b}bb'\right)\right].
  \end{align}
  Let $\hQs=\Qs(\hat{\beta},\hat{\Omega})$. If $m_{\max}\leq \Ni/n$, then the
  Hessian, evaluated at $(\hat{\beta},\hat{\lambda},\hat{\Omega})$, is singular.
  Otherwise, it is given by:
  \begin{equation*}
    \mathcal{H}_{\re}(\hat\beta,\hat\lambda,\hat\Omega)=
    \begin{pmatrix}
      \frac{d(\hat{\lambda})}{\hat{b}'\hat{\Omega} \hat b}
      (\hQs\hat{\Omega}_{22}
      - T_{22}) & 0 & \hat{\mathcal{H}}_{1, 3: 5}\\
      0&-\frac{1}{2}\frac{\Ni }{(\Ni / n + \hat\lambda)^{2}}
      & \hat{\mathcal{H}}_{2, 3: 5}\\
      \hat{\mathcal{H}}_{1,3:5}'&\hat{\mathcal{H}}_{2,3:5}' &
      \hat{\mathcal{H}}_{3: 5, 3: 5}
    \end{pmatrix},
  \end{equation*}
  where
  \begin{align*}
    \hat{\mathcal{H}}_{1, 3: 5}
    &= 
      \frac{1}{2} \frac{d(\hat{\lambda})\hQs}{
      \hat b'\hat \Omega \hat{b}}
      \left(2\frac{e_{2}'\hat \Omega \hat b}{\hat b'\hat \Omega \hat b}
      \hat{b}\otimes \hat{b}-\hat{b}\otimes e_{2}-e_{2}\otimes b \right)'D_{2},\\
    \hat{\mathcal{H}}_{2,3:5}
    &=\frac{1}{2}\frac{\Ni}{(\Ni / n +
      \hat\lambda)^{2}}\left( \frac{\hQs
      }{\hat b\hat \Omega\hat b}\hat b\otimes
      \hat{b} -\mkvec(\hat\Omega^{-1} T \hat\Omega^{-1})\right)'D_{2}
      =-\frac{1}{2}\frac{\Ni}{(\Ni / n +
      \hat\lambda)^{2}}\frac{m_{\max}}{\hat{a}S^{-1}\hat{a}}\left(
      \hat{\Omega}^{-1}\hat{a}\otimes
      \hat{\Omega}^{-1}\hat{a}
      \right)'D_{2},
    \\
    \hat{\mathcal{H}}_{3:5,3:5}
    & =-\frac{(n - \Ne
      )}{2}D_{2}'\left(\left(\hat\Omega^{ - 1}-\frac{\hat c\hat b\hat b'}{
      \hat{b}'\hat\Omega\hat{b}}\right)\otimes\left(\hat\Omega^{ - 1}-\frac{
      \hat{c}\hat{b}\hat{b}'}{\hat b'\hat\Omega\hat b}\right) - (2\hat{c}-
      \hat{c}^{2})\frac{\hat{b}\hat b'}{\hat b'\hat\Omega\hat b}\otimes \frac{
      \hat{b}\hat b'}{\hat b'\hat\Omega\hat b} \right)D_{2}.
  \end{align*}
  By the formula for block inverses, the upper $2\times 2$ submatrix of the
  inverse Hessian is given by:
  \begin{equation}\label{eq:blockinverse}
    \hat{\mathcal{H}}^{{1:2,1:2}}(\hat\beta,\hat\lambda,\hat\Omega)=
    \left(\hat{\mathcal{H}}_{1:2,1:2}-\hat{\mathcal{H}}_{1:2,3:5}
      \hat{\mathcal{H}}_{3:5,3:5}^{-1}\hat{\mathcal{H}}_{1:2,3:5}'\right)^{-1}.
  \end{equation}
  Applying \Cref{th:kronecker} and using the fact that
  $N_{d}(A\otimes A)=N_{d}(A\otimes A)N_{d}=(A\otimes A)N_{d}$ \citep[Lemma
  2.1(v)]{mn80} yields:
  \begin{equation*}
    \hat{\mathcal{H}}_{3:5,3:5}^{-1}=
    -\frac{2}{n-\Ne}L_{2}N_{2}\left[  \left(\hat\Omega+\frac{\hat c}{1-\hat c}
        \frac{\hat\Omega\hat b\hat b'\hat\Omega }{
          \hat{b}'
          \hat\Omega\hat b}\right)
      \otimes   \left(\hat\Omega+\frac{\hat c}{1-\hat c}
        \frac{\hat\Omega\hat b\hat b'\hat\Omega }{\hat{b}'
          \hat\Omega\hat b}\right)+\frac{\hat c^{2}-2\hat c}{(1-\hat c)^{2}}
      \frac{  \hat\Omega\hat b    \hat b'\hat\Omega \otimes \hat\Omega\hat b
        \hat b'\hat\Omega  }{(\hat b'\hat\Omega\hat b)^{2}}\right] N_{2} L_{2}',
  \end{equation*}
  It follows that
  \begin{equation*}
    \hat{\mathcal{H}}_{1,3:5}\hat{\mathcal{H}}_{3:5,3:5}^{-1}
    \hat{\mathcal{H}}_{1,3:5}'
    =-\frac{(n-\Ne)\hat c^{2}}{1-\hat c}\frac{\abs{\Omega}}{(b'\Omega b)^{2}}.
  \end{equation*}
  Finally, since
  $ \hat{\mathcal{H}}_{2,3:5}\hat{\mathcal{H}}_{3:5,3:5}^{-1}
  \hat{\mathcal{H}}_{1,3:5}'=0$, \Cref{eq:blockinverse} combined with the
  expression in the previous display yields
  \begin{align*}
    \h{\mathcal{H}}{re}^{11}
    &=\left(\hat{\mathcal{H}}_{11}-
      \hat{\mathcal{H}}_{1,3:5}\hat{\mathcal{H}}_{3:5,3:5}^{-1}
      \hat{\mathcal{H}}_{1,3:5}'\right)^{-1}=\frac{\hat b'\hat \Omega\hat{b}
      (\hat\lambda+\Ni/n)}{n\hat\lambda}\left(
      \hQs\hat\Omega_{22}-T_{22}+\frac{\hat c}{1-\hat c}\frac{\hQs}{\hat a'
      \hat{\Omega}^{-1}\hat{a}} \right)^{-1},
  \end{align*}
  which yields the result.

  It remains to show that the inverse Hessian is consistent for $V_{\liml, N}$.
  To this end, note that
  $m_{\min}=\frac{\h{b}{liml}'T\h{b}{liml}}{\h{b}{liml}'S \h{b}{liml}}\inprob
  \alphai$ by \Cref{th:st:adist:valid} and consistency of $\hat{\beta}$. By
  continuity of the trace operator, and \Cref{th:st:adist:valid}
  \begin{equation*}
    m_{\max}=\trace(S^{-1}T)-m_{\min}=\trace(S^{-1}T)
    -\frac{\h{b}{liml}'T\h{b}{liml}}{\h{b}{liml}'S
      \h{b}{liml}}
    \inprob 2\alphai+\lambda-\alphai=\lambda+\alphai.
  \end{equation*}
  Consistency of $\hat{\Omega}$ then follows by consistency of $\hat{\lambda}$
  and $\hat{\beta}$, \Cref{th:st:adist:valid}, and Slutsky's Theorem. It also
  follows that
  \begin{equation*}
    \hQs=(n-\Ne)\frac{\hat b'T\hat b}{\hat b'\tilde{S} \hat b}=
    \frac{    (n-\Ne)\hat b'T\hat b}{(\nus)\hat b'S \hat b+n  b'T
      \hat{b}}
    =\left(\frac{1}{1-\Ne/n}+\frac{\nus}{(n-\Ne)m_{\min}}\right)^{-1}
    \inprob \alphai.
  \end{equation*}
  Hence,
  \begin{equation*}
    \frac{\hat c}{1-\hat c}\inprob \frac{\alphai\lambda}{\alphai(1-\alphae)+
      (\alphanu)\lambda},
  \end{equation*}
  so that
  \begin{equation*}
    \begin{split}
      -n\h{H}{re}^{11}&\inprob- \frac{b'\Omega
        b(\alpha_{K}+\lambda)}{\lambda}\left( -\frac{\lambda}{a'\Omega^{-1}
          a}+\frac{\lambda
          \alpha_{K}^{2}}{a'\Omega^{-1}a\left((1-\alpha_{K}-\alphae
            )\lambda+(1-\alphae )\alpha_{K}\right)}
      \right)^{-1}\\
      &=\frac{b'\Omega b a'\Omega^{-1}a}{\lambda^{2}}\left(
        \lambda+\frac{(1-\alphae )\alpha_{K}}{1-\alphae
          -\alpha_{K}}\right)=\mathcal{V}_{\liml,N},
    \end{split}
  \end{equation*}
  which completes the proof.
\end{sproof}

\begin{sproof}[\Cref{th:md-re:equivalence2}]
  The objective function evaluates as:
  \begin{equation}\label{eq:md:Wre:objective}
    {\mathcal{Q}}_{n}(\beta,\Xi_{22,n};\h{W}{re})
    =\trace((TS^{-1}-(\Ni/n)I_{2})^{2})
    +\Xi_{22,n}\cdot a'S^{-1}a\left[
      \Xi_{22,n}\cdot a'S^{-1}a-2 \Qt(\beta,S)+2\Ni/n
    \right].
  \end{equation}
  Consider first minimizing the objective function with respect to $\Xi_{22,n}$,
  holding $\beta$ fixed. Let $\hat{\Xi}_{\beta}$ denote the resulting estimator.
  Since the derivative
  $\partial{\mathcal{Q}}_{n}(\beta,\Xi_{22,n};\h{W}{re})/\partial \Xi_{22,n}$
  equals zero at $\Xi_{22,n}=(\Qt(\beta,S)-\Ni/n)/(a'S^{-1}a)$ and is positive
  for $\Xi_{22,n}\geq (\Qt(\beta,S)-\Ni/n)/(a'S^{-1}a)$, we get
  \begin{equation}\label{eq:Xi:md}
    \hat\Xi_{\beta}=\frac{\max\{\Qt(\beta,S)-\Ni/n,0\}}{a'S^{-1}a}.
  \end{equation}
  Therefore, the objective function with $\Xi_{22,n}$ concentrated out is given by
  \begin{equation*}
    {\mathcal{Q}}_{n}(\beta,\hat{\Xi}_{\beta})
    =\trace((TS^{-1}-(\Ni/n)I_{2})^{2})
    -(\Qt(\beta,S)-\Ni/n)^{2}\cdot \mathbb{1}\{\Qt(\beta,S)\geq\Ni/n\},
  \end{equation*}
  where $\mathbb{1}\{\cdot \}$ denotes the indicator function. Since
  $\max_{\beta}\Qt(\beta,S)=m_{\max}$, with the maximum attained at
  $\h{\beta}{liml}$, it follows that if $m_{\max}> \Ni/n$, the objective
  function~\eqref{eq:md:Wre:objective} is uniquely minimized at
  $(\h{\beta}{re},\h{\lambda}{re}/\h{a}{re}'S^{-1}\h{a}{re})$. If
  $m_{\max}\leq \Ni/n$, then $ {\mathcal{Q}}_{n}(\beta,\Xi_{22,n};\h{W}{re})$ is
  minimized at
  ${\Xi}_{22,n}=0=\h{\lambda}{re}/\h{a}{re}'\h{\Omega}{re}^{-1}\h{a}{re}$ and an
  arbitrary $\hat{\beta}$, so that in particular we can set
  $\h{\beta}{re}=\beta$. Therefore, Part~\ref{item:md-re-1} of
  \Cref{th:md-re:equivalence2} follows if we can show that if $m_{\max}>\Ni/n$,
  then $\h{a}{re}'S^{-1}\h{a}{re}=\h{a}{re}'\h{\Omega}{re}^{-1}\h{a}{re}$. Using
  the notation $\tilde{S}=(\nus)S+nT$, we have
  \begin{equation}\label{eq:aoa:asa}
    \begin{split}
      \h{a}{re}'\h{\Omega}{re}^{-1}\h{a}{re}&= (n-\Ne)\h{a}{re}'\left(\tilde{S}
        - n\frac{m_{\max}-\Ni/n}{\h{a}{re}'S^{-1}\h{a}{re} }
        \h{a}{re}\h{a}{re}'\right)^{-1}\h{a}{re}\\
      &=-(n-\Ne)\frac{
        \h{a}{re}'\tilde{S}^{-1}\h{a}{re}\h{a}{re}'S^{-1}\h{a}{re}}{
        n(m_{\max}-\Ni/n)\h{a}{re}\tilde{S}^{-1}\h{a}{re}-\h{a}{re}'S^{-1}\h{a}{re} }\\
      &=-(n-\Ne)\left( nm_{\max}-\Ni-
        \frac{\abs{\tilde{S}}}{\abs{S}}\frac{\h{b}{re}'S\h{b}{re}}{\h{b}{re}\tilde{S}\h{b}{re}}
      \right)^{-1}\h{a}{re}'S^{-1}\h{a}{re}\\
      &=\h{a}{re}'S^{-1}\h{a}{re},
    \end{split}
  \end{equation}
  where the first line follows from the definition of $\h{\Omega}{re}$ and
  $\h{\lambda}{re}$ given in \Cref{th:re-theorem}, the second line follows by
  the Woodbury identity, the third line follows from \Cref{eq:equalities:aoa},
  and the fourth line follows from \Cref{eq:equalities:yy}.

  To prove the second part of~\Cref{th:md-re:equivalence2}, I show that whenever
  the weight matrix satisfies
  \begin{align*}
    \hat W_{n}\inprob
    & \overbar{c} D_{2}'\Phi_{t}^{-1}D_{2},
    &\text{where}&
    &      \Phi_{t}
    &= \Omega\otimes \Omega+\Omega\otimes t mm' +t mm'\otimes\Omega,
  \end{align*}
  for some constants $\overbar{c}>0$ and $t\geq 0$, with $m=\Xi_{22}^{1/2}a$,
  then it is asymptotically optimal. Since we can write
  $\Phi_{t}=((\Omega+tmm')\otimes (\Omega+tmm')-t^{2}(mm')\otimes (mm'))$, by
  \Cref{th:kronecker}, and the identity $\lambda=m'\Omega^{-1}m$,
  \begin{equation*}
    \begin{split}
      \Phi_{t}^{-1} &=(\Omega^{-1}\otimes \Omega^{-1})\left[\left(\Omega-
          \frac{t mm'}{1+t\lambda}\right) \otimes \left(\Omega-\frac{tmm'}{1+
            t\lambda}\right)+ \frac{t^{2}(mm')\otimes
          (mm')}{(1+2t\lambda)(1+t\lambda)^{2}}\right](\Omega^{-1}\otimes
      \Omega^{-1}).
    \end{split}
  \end{equation*}
  By \Cref{th:st:adist:valid}, the asymptotic variance of the moment condition
  \begin{equation}
    \vech(T-(\Ni/n)S-\Xi_{22,n}aa')\label{eq:moment:Xi}
  \end{equation}
  is given by
  \begin{equation}\label{eq:avar:g}
    \Delta= 2L_{2}N_{2}\left[\tau \Omega\otimes \Omega+\Omega\otimes (mm')
      +(mm')\otimes\Omega\right] L_{2}',
  \end{equation}
  where $\tau=\alphai(1-\alphae)/(1-\alphai-\alphae)$. Suppose first that
  $\tau>0$. Then $\Delta$ is invertible, and by
  \Cref{th:kronecker}\ref{it:kronecker-lnd}, its inverse is given by
  $\Delta^{-1}=\frac{1}{2\tau}D_{2}'\Phi_{1/\tau}^{-1}D_{2}$. A necessary and
  sufficient condition for optimality is that for some matrix $C_{t}$
  \citep[Section 5.2]{nm94},
  \begin{equation}\label{eq:efficient:misspec}
    (D_{2}'\Phi^{-1}_{t}D_{2})G
    =\Delta^{-1}GC_{t}=\frac{1}{2\tau}D_{2}'\Phi_{1/\tau}^{-1}D_{2}GC_{t},
  \end{equation}
  where $G$ is the derivative of the moment condition~\eqref{eq:moment:Xi},
  given by:
  \begin{equation*}
    G=-L_{2}M,\qquad
    M=    \begin{pmatrix}
      \Xi_{22}^{1/2}(m\otimes e_{1}+e_{1}\otimes m)&
      \frac{1}{\Xi_{22}}m\otimes m
    \end{pmatrix},
  \end{equation*}
  where $e_{1}=(1,0)'$. Since for a symmetric matrix
  $A\in\mathbb{R}^{2\times 2}$, $D_{2}L_{2}\mkvec(A)=\mkvec(A)$
  \citep[p.~427]{mn80}, it follows that $D_{2}G=-M$, so that
  \begin{equation*}
    \Phi_{t}^{-1}D_{2}G=
    -(\Omega^{-1}\otimes \Omega^{-1})    \begin{pmatrix}
      \frac{\Xi_{22}^{1/2}}{1+t\lambda}\left(
        m\otimes e_{1}+e_{1}\otimes m-\frac{2t
          m'\Omega^{-1}e_{1}}{1+2t\lambda}m\otimes m
\right)
      & \frac{1}{\Xi_{22}(1+2t\lambda)}m\otimes m
    \end{pmatrix}.
  \end{equation*}
  It then follows that~\eqref{eq:efficient:misspec} holds with
  \begin{equation*}
    C_{t}=2\tau
    \begin{pmatrix}
      \frac{1+\lambda/\tau}{1+\lambda t}&0\\
      \frac{2m'\Omega^{-1}e_{1}\Xi^{3/2}}{1+t\lambda}
      \left(\frac{1}{\tau}-t\frac{1+2\lambda/\tau}{1+2\lambda t}\right) &
      \frac{1+2\lambda/\tau}{1+2 \lambda t}
    \end{pmatrix}.
  \end{equation*}
  If $\tau=0$, then the asymptotic variance $\Delta$ given in \Cref{eq:avar:g}
  is degenerate, since one of the three moment conditions given in
  \Cref{eq:moment:Xi} is asymptotically redundant: the first moment condition
  equals $2\beta$ times the second minus $\beta^{2}$ times the third. In this
  case, any weight matrix that puts positive weight on at least two of the
  moment conditions will be optimal, and in particular $\hat{W}_{n}$ is optimal.
\end{sproof}
\begin{sproof}[\Cref{th:moment-variance}]
  Part~\ref{it:moment-variance} of the Lemma follows from
  \Cref{th:moment-distro}. Consistency of $\hat{\Psi}_{3}$ and $\hat{\Psi}_{4}$
  follows from Lemma A.7 in \citet{anatolyev13}. Finally, since $ZZ'$ is a
  projection matrix, $\norm{ZZ'\diag(H)}\leq \norm{\diag(H)}$, so that
  \begin{equation*}
    \var(\hat{\mu})=\frac{1}{n\Ni}\Omega_{22}\norm{ZZ'\diag(H)}^{2}
    \leq\frac{1}{n\Ni}\Omega_{22}\diag(H)'\diag(H)=\frac{\Omega_{22}}{n}
    \delta(1+o(1))=o(1).
  \end{equation*}
  Thus, $\hat{\mu}\inprob \mu$ by Markov inequality.
\end{sproof}
\begin{sproof}[\Cref{th:umd}]
  It follows by \Cref{th:moment-distro} and the Delta method that
  \begin{equation*}
    \sqrt{n}\left(\h{\beta}{umd}-\Xi_{12,n}/\Xi_{22,n}\right)
    \indist \mathcal{N}(0,V_{\umd}),
  \end{equation*}
  where, letting $A=(e_{2}b'+be_{2}')/2\Xi_{22}$, $e_{2}=(0,1)'$,
  \begin{equation*}
    \begin{split}
      V_{\umd}&=\mkvec(A)'\Delta\mkvec(A)\\
      &=\trace(4A \Omega A\Xi+2\tau A\Omega A \Omega) +\alphai
      \delta\left(\E[(v_{i}' A v_{i})^{2}]-\trace(\Omega A)^{2}-2\trace(A\Omega
        A \Omega)\right) +4\alphai^{1/2}
      (\tilde{\mu}+\mu \beta,\mu)\E[ A v_{i}v_{i}' Av_{i}]\\
      &=V_{\liml,N}+\frac{2\tau(e_{2}'\Omega b)^{2}+2\alphai^{1/2}
        {\mu}\E[v_{2i}\epsilon_{i}^{2}] +
        \alphai\delta(\E[v_{2i}^{2}\epsilon_{i}^{2}]-3(e_{2}'\Omega b)^{2}
        -\abs{\Omega})}{\Xi_{22}^{2}}+\frac{\Omega_{22}\abs{\Xi}/\Xi_{22}
        +2\alphai^{1/2}\tilde{\mu}\E[v_{2i}^{2}\epsilon_{i}]}{\Xi_{22}^{2}}\\
      &= V_{\liml}+\frac{(2\tau+\alphai\delta\kappa)(e_{2}'\Omega b)^{2}
        +2\gamma\left(\alphai\delta \E[(v_{2i}-\gamma
          \epsilon_{i})\epsilon_{i}^{3}] +
          \alphai^{1/2}\mu\E[\epsilon_{i}^{3}]\right)
      }{\Xi_{22}^{2}}+\frac{\Omega_{22}\abs{\Xi}/\Xi_{22}
        +2\alphai^{1/2}\tilde{\mu}\E[v_{2i}^{2}\epsilon_{i}]}{\Xi_{22}^{2}},
    \end{split}
  \end{equation*}
  from which the result follows.
\end{sproof}
\begin{sproof}[\Cref{th:overid}]
  We have:
  \begin{equation*}
    \begin{split}
      (m_{\min}-\alphai)&= \frac{\h{b}{liml}'
        (T-\alphai S)\h{b}{liml}}{\h{b}{liml}'S\h{b}{liml}}\\
      &= \frac{\h{b}{liml}'(T-\alphai
        S-\lambda_{n}aa'/(a'\Omega^{-1}a))\h{b}{liml}}{\h{b}{liml}'S\h{b}{liml}}+
      \frac{\lambda_{n}
        (a'\h{b}{liml})^{2}}{(a'\Omega^{-1}a)\h{b}{liml}'S\h{b}{liml}}\\
      &= \frac{(\h{b}{liml}\otimes \h{b}{liml})'\mkvec\left(T-\alphai
          S-\lambda_{n}aa'/(a'\Omega^{-1}a)\right)}{\h{b}{liml}'S\h{b}{liml}}
      +\frac{\lambda_{n}(\h{b}{liml}-\beta)^{2}}{(a'\Omega^{-1}a)
        \h{b}{liml}'S\h{b}{liml}}\\
      &= \frac{(\h{b}{liml}\otimes \h{b}{liml})' \mkvec\left(T-(\Ni/n)
          S-\Xi_{22,n}aa'\right)}{\h{b}{liml}'S\h{b}{liml}}+O_{p}(n^{-1})\\
      &= \frac{(b\otimes b)' \mkvec\left(T-(\Ni/n)
          S-\Xi_{22,n}aa'\right)}{b'\Omega b}+O_{p}(n^{-1}),
    \end{split}
  \end{equation*}
  where the first line follows from the identity
  $m_{\min}=\Qs(\h{\beta}{liml},S)$, the second and third line follows by
  algebra, the fourth line and the last line follow from $\sqrt{n}$-rate of
  convergence $\h{\beta}{liml}$ and $T-(\Ni/n)S$. Expanding the numerator then
  yields
  \begin{equation*}
    \frac{n}{\sqrt{\Ni}}(m_{\min}-\alphai)=\frac{(b\otimes b)'\mkvec(
      V'(H/\sqrt{\Ni}) V)}{b'\Omega b}+O_{p}(\Ni^{-1/2})=
    \frac{  \epsilon'(H/\sqrt{\Ni}) \epsilon}{b'\Omega b}+O_{p}(\Ni^{-1/2}).
  \end{equation*}
  If $\Ni\to\infty$, then by \Cref{th:clt,th:lemma-p4}, with $\Pm=H/\sqrt{\Ni}$,
  \begin{equation*}
    \frac{n}{\sqrt{\Ni}}(m_{\min}-\alphai)\indist \mathcal{N}
    (0, 2(1-\alphae)/(\alphanu)+\delta\kappa).
  \end{equation*}
  If $\Ni=K$ is fixed, then
  $\mkvec(\hat{\Pi}-\Pi)=\sum_{i}v_{i}\otimes z_{i} \indist\mathcal{N}(0,\Omega
  \cdot I_{K})$, since the Lyapunov condition is implied by
  $\sum_{i=1}^{n}\norm{z_{i}}^{2+\nu}=\sum_{i}(ZZ')_{ii}^{1+\nu/2}\leq
  \max_{i}(ZZ')_{ii}^{\nu/2}\trace(ZZ')=o(1)$ for any $\nu>0$. It then follows
  by standard arguments that $nm_{\min}\indist \chi^{2}_{K-1}$, which proves the
  first part.

  To prove the second part, I use the approximation from \citet{peiser43} (see
  also \citealp{ag11}) that as $k\to \infty$,
  \begin{equation*}
    q_{1-\ns}^{\chi^{2}_{k}}=k+\Phi^{-1}(1-\ns)\sqrt{2k}+O(1),
  \end{equation*}
  where $q_{1-\ns}^{\chi^{2}_{k}}$ denotes the $1-\ns$ quantile of a $\chi^{2}$
  distribution with $k$ degrees of freedom. Therefore, if $\Ni\to\infty$,
  letting
  $c=\Phi(\sqrt{\frac{(1-\alphai)}{\alphanu}+
    \frac{\delta\kappa}{2}}\Phi^{-1}(\ns))$
  \begin{equation*}
    \begin{split}
      \Prob\left(nm_{\min}\geq q^{\chi^{2}_{\Ni-1}}_{1-c}\right)&=
      \Prob\left(nm_{\min}/\sqrt{\Ni}\geq
        \sqrt{\Ni}+\Phi^{-1}(1-c)\sqrt{2}+o(1)\right)\\
      &= \Prob\left(nm_{\min}/\sqrt{\Ni}-\sqrt{\Ni}\geq
        \Phi^{-1}(1-c)\sqrt{2}+o(1)\right)\\
      &= \Prob\left(\mathcal{N}(0,1)+o_{p}(1)\geq \Phi^{-1}(1-c)\sqrt{
          \frac{2(\alphanu)}{2(1-\alphae)+(\alphanu)\delta\kappa}}+o(1)\right)\\
      &= \Phi\left(\Phi^{-1}(c)\sqrt{
          \frac{2(\alphanu)}{2(1-\alphae)+(\alphanu)\delta\kappa}}\right)+o(1)\\
      &=\ns+o(1).
    \end{split}
  \end{equation*}
  If $\Ni$ is fixed, then, since
  $\sum_{i=1}^{n}(ZZ')_{ii}^{2}\leq \max_{j}(ZZ')_{jj}\trace(ZZ')$, it follows
  that $\delta_{n}=\sum_{i=1}^{n}(ZZ')_{ii}^{2}/\Ni+o(1)=o(1)$. Thus,
  $c=\ns+o(1)$, and so
  $ \Prob\left(nm_{\min}\geq q^{\chi^{2}_{\Ni-1}}_{1-c}\right)=\ns+o(1)$. Since
  $\delta_{n}$ and $\hat{\kappa}$ are consistent estimators of $\delta$ and
  $\kappa$, the assertion of the theorem follows.
\end{sproof}
\end{appendices}

\bibliographystyle{ecca}%
\renewcommand{\bibfont}{\footnotesize}%
\bibliography{many-iv-library}
\end{document}


\maketitle

\section{Additional details for calculations in main text}
Let $e_{1}=(1,0)'$ and let $e_{2}=(0,1)'$.

\subsection{Additional details for
  \texorpdfstring{\Cref{sec:suff-stat-limit}}{Section 2.2}}
This section shows that the block of the inverse information matrix based on the
limited information likelihood corresponding to $\beta$ is given by
$n^{-1} b'\Omega b\cdot a'\Omega^{-1}a/\lambdan$.

The distribution of the statistics $\hat{\Pi}$ and $S$ is given by
\begin{align}
  \mkvec(\hat\Pi)&\sim \mathcal{N}_{2\Ni}\left((a'\Omega^{-1}a/n)^{-1/2} a\otimes
                   \etan,\Omega\otimes
                   I_{\Ni}\right)\label{eq:s:hatpi1:distro},\\
  (\nus)S&\sim \mathcal{W}_{2}(\nus, \Omega),\label{eq:s:ymy:distro}
\end{align}
with $\hat\Pi$ independent of $S$, where $\mathcal{W}_{2}(\nus, \Omega)$ denotes
a Wishart distribution with $\nus$ degrees of freedom, and scale matrix
$\Omega$. Their densities are therefore given by
\begin{equation}\label{eq:re:ymy:density}
  \begin{split}
    f_{\hat\Pi}(\hat\Pi;\beta, \etan, \Omega)
    &=\frac{\abs{\Omega}^{-\Ni/2}}{(2\pi)^{\Ni}} \exp\left(-\frac{n}{2}
      \left(\trace(\Omega^{-1}T)+\etan'\etan-2
        \frac{\etan'\hat\Pi\Omega^{-1}a}{(na'\Omega^{-1}a)^{1/2}}
      \right)\right),\\
    f_{S}(S;\Omega) &=
    C_{\nu} \abs{S}^{(\nu-3)/2} \abs{\Omega}^{-\nu/2}
    e^{-\frac{\nu}{2}\trace(\Omega^{-1} S)},
  \end{split}
\end{equation}
where $\nu=\nus$, and
$C_{\nu}^{-1}= (2/\nu)^{\nu}\pi ^{1/2}\Gamma (\nu/2)\Gamma ((\nu-1)/2)$, with
$\Gamma$ denoting the gamma function.

It follows that the limited information likelihood is given by
\begin{equation}\label{eq:liml:likelihood}
  \mathcal{L}_{\li,n}(\beta,\etan,\Omega)=
  \frac{C_{\nu}\abs{S}^{(\nu-3)/2}}{(2\pi)^{\Ni}}\cdot
  \abs{\Omega}^{-(n-\Ne)/2}
  e^{-\frac{1}{2}\left(
      \trace(\Omega^{-1}\tilde{S})
      +n\etan'\etan-2n\frac{\etan'\hat\Pi\Omega^{-1}a
      }{(na'\Omega^{-1}a)^{1/2}}
    \right)},
\end{equation}
where $\tilde{S}=nT+\nu S$.

The score is given by
\begin{equation*}
  \begin{split}
    \mathcal{S}_{\beta}(\beta, \etan, \Omega)&=n^{1/2} \frac{\etan'
      \hat\Pi\Omega^{-1}d(\beta, \Omega)}{(a\Omega^{-1}a)^{1/2}},
    \\
    \mathcal{S}_{\etan}(\beta, \etan, \Omega)&= n\left(\frac{n^{-1/2}\hat\Pi
        \Omega^{-1}a}{(a'\Omega^{-1}a)^{1/2}}-\etan\right),\\
    \mathcal{S}_{\vech(\Omega)}(\beta, \etan, \Omega)&=\frac{1}{2}\tilde D'
    \left[\mkvec\left(\tilde{S} -(n-\Ne)\Omega
      \right)-\frac{2n^{1/2}}{(a'\Omega^{-1}a)^{1/2}} \hat\Pi'\etan \otimes a
      +\frac{n^{1/2}\etan'\hat\Pi\Omega^{-1}a}{(a'\Omega^{-1}a)^{3/2}} a\otimes
      a\right],
  \end{split}
\end{equation*}
where $\tilde D=(\Omega^{-1}\otimes \Omega^{-1})D_{2}$, $D_{2}$ is the
duplication matrix, and
$d(\Omega, \beta)=e_{1}-a \frac{a'\Omega^{-1}e_{1}}{a'\Omega^{-1}a}$. Let
$\mathcal{H}_{\beta\etan}(\beta, \etan, \Omega)$ denote the $\beta$-$\etan$ block
of the Hessian, and similarly for the other blocks. By taking derivatives of the
score, we obtain that
\begin{align*}
  \mathcal{H}_{\beta\beta}(\beta, \etan, \Omega)
  &=
    -2S_{\beta}(\beta, \etan, \Omega)
    \frac{a'\Omega^{-1}e_{1}}{a'\Omega^{-1}a}-\frac{n^{1/2}
    \etan'\hat{\Pi}\Omega^{-1}a}{(a'\Omega^{-1}a)^{1/2}}\frac{
    1
    }{b'\Omega b\cdot a'\Omega^{-1}a},\\
  \mathcal{H}_{\etan\beta}(\beta, \etan, \Omega)
  &=\frac{n^{1/2}\hat{\Pi}\Omega^{-1}d(\Omega, \beta)}{(a\Omega^{-1}a)^{1/2}},
  \\
  \mathcal{H}_{\vech(\Omega)\beta}(\beta, \etan, \Omega)
  &=
    \tilde{D}'\left(
    \frac{\mathcal{S}_{\beta}(\beta, \etan, \Omega)}{2 a'\Omega^{-1}a}
    a\otimes a
    +\frac{n^{1/2}}{\sqrt{a'\Omega^{-1}a}}
    \left(\frac{\etan'\hat{\Pi}\Omega^{-1}a}{a'\Omega^{-1}a}
    a
    -2\hat{\Pi}'\etan
    \right)\otimes d(\Omega, \beta)
    \right),
\end{align*}
where we use the identities
$e_{1}'\Omega^{-1}e_{1}a'\Omega^{-1}a-
(a'\Omega^{-1}e_{1})^{2}=\abs{\Omega}^{-1}$,
$D_{2}(v_{1}\otimes v_{2})=D_{2}(v_{2}\otimes v_{1})$ for any vectors
$v_{1}, v_{2}$, and
$(\hat{\Pi}'\etan)\otimes a=\mkvec(a\etan'\hat{\Pi}) =(\hat{\Pi}'\otimes
a)\etan$. Since $a'\Omega^{-1}d(\Omega, \beta)=0$, it follows that
$\E[\mathcal{H}_{\vech(\Omega)\beta}(\beta, \etan, \Omega)]=0$ and
$\E[\mathcal{H}_{\etan\beta}(\beta, \etan, \Omega)]=0$. Thus, the (1,1) element
block of the inverse information matrix is given by
\begin{equation*}
  \mathcal{I}_{\li, n}^{11}(\beta, \etan, \Omega)=
  -\frac{1}{E[\mathcal{H}_{\beta\beta}(\beta, \etan, \Omega)]}=
  \frac{a'\Omega^{-1}a b'\Omega b}{n\etan'\etan}=
  \frac{a'\Omega^{-1}a b'\Omega b}{n\lambdan}
  ,
\end{equation*}
as stated in the main text.

\subsection{Additional details for
  \texorpdfstring{\Cref{sec:effic-minim-dist}}{Section 4.2}}
Consider the groups example, so that $z_{ij}^{*}=1$ if individual $i$ belongs to
group $j$ and zero otherwise, and let $W=\iota_{n}$, where $\iota_{n}$ denotes
to an $n$-vector of ones, so that $(W'(W'W)W')_{ij}=1/n$. Let $\nu$ denote a
$\Ni$-vector with elements $\nu_{j}=n_{j}$, and let $\diag(\nu)$ denote a
diagonal matrix with elements $\nu_{j}=n_{j}$ on the diagonal. It then follows
that $\tilde{Z}=Z^{*}-W'(W'W)W'Z^{*}=Z^{*}-\iota_{n}\nu'/n$,
$\tilde{Z}'\tilde{Z}=\diag(\nu)-\nu\nu'/n$,
$(\tilde{Z}'\tilde{Z})^{-1}=\diag(\nu)^{-1}+\iota_{\Ni}\iota_{\Ni}'/n$, and
\begin{equation*}
  (ZZ')_{ii}=(\tilde{Z}(\tilde{Z}'\tilde{Z})^{-1}\tilde{Z}')_{ii}
  =1/n_{j(i)}-1/n,
\end{equation*}
where ${j(i)}$ denotes the group index that individual $i$ belongs to. Since
$(W'(W'W)W)_{ii}=1/n$, it follows that
\begin{equation*}
  H_{ii}=(ZZ')_{ii}-\frac{\Ni}{n-1-\Ni}\left(1-(ZZ)_{ii}-1/n\right)
  =\frac{n-1}{n-1-\Ni}\left(
    \frac{1}{n_{g(i)}}- \frac{1+\Ni}{n}\right)
\end{equation*}
Consequently,
\begin{multline*}
  \delta_{n}= \diag(H)'\diag(H)/\Ni =\frac{(n-1)^{2}}{\Ni(n-1-\Ni)^{2}}
  \sum_{j=0}^{\Ni}n_{j}\left(
    \frac{1}{n_{j}}- \frac{1+\Ni}{n}\right)^{2}\\
  =\frac{(n-1)^{2}}{(n-1-\Ni)^{2}\Ni}
  \left(\sum_{j}\frac{1}{n_{j}}-\frac{(\Ni+1)^{2}}{n}\right).
\end{multline*}

\subsection{Additional details for
  \texorpdfstring{\Cref{sec:unrestr-minim-dist}}{Section 5.2}}\label{sec:addit-deta-crefs}
I illustrate the minimization of the minimum distance objective function given
in \Cref{eq:md:ure} in the paper subject to the constraint
$\Xi_{11,n}\geq \beta^{2}\Xi_{22,n}$. For concreteness and simplicity, consider
the random-effects weight matrix $\h{W}{re}=D_{2}'(S^{-1}\otimes S^{-1})D_{2}$,
and suppose that the errors are normal. The solution is given by
\begin{equation*}
  \begin{pmatrix}
    \hat{\Xi}_{11}&\hat{\Xi}_{22} &\hat{\beta}
  \end{pmatrix}=\begin{cases}
    \begin{pmatrix}
      \h{\Xi}{11,umd}&\h{\Xi}{22,umd}& \h{\beta}{umd}
    \end{pmatrix}    & \text{if $S-(\Ni/n)T$ is positive semi-definite,} \\
    \begin{pmatrix}
      \h{\Xi}{22,re}\h{\beta}{re}^{2}&\h{\Xi}{22,re}& \h{\beta}{re}
    \end{pmatrix} & \text{otherwise.}
  \end{cases}
\end{equation*}
When \Cref{an:pr} does not hold, then $T-(\Ni/n)S$ will be positive definite
with probability approaching one so that the restriction will not bind
asymptotically. Otherwise, under \Cref{an:n,an:mi}, its distribution is given by
\begin{equation}\label{eq:ure:boundary-distro}
  \sqrt{n}\left(\hat{\beta}-\beta\right)\indist \sqrt{V_{\liml,N} }\mathcal{Z}_{2}+
  \frac{ \sqrt{2\tau}(b'\Omega e_{2})}{\Xi_{22}} \max(\mathcal{Z}_{1},0),\qquad
  \begin{pmatrix}
    \mathcal{Z}_{1}\\\mathcal{Z}_{2}
  \end{pmatrix}
  \sim
  \mathcal{N}_{2}(0,I_{2}),
\end{equation}
where $V_{\liml,N}$ is given in \Cref{eq:avar:liml} in the paper and
$\tau= \frac{\alphai(1-\alphae)}{1-\alphai-\alphae}$. This result follows by
verifying the conditions for Theorem 1 in \citet{andrews02}.
The asymptotic distribution is non-standard, and since
$\E \max(\mathcal{Z}_{1},0)>0$, $\hat{\beta}$ is asymptotically biased. Recall
that for \umd,
\begin{equation}\label{eq:mbtsls:wi}
  \sqrt{n}\left(\h{\beta}{umd}-\beta\right)\indist
  \sqrt{V_{\liml} }\mathcal{Z}_{2}+
  \frac{ \sqrt{2\tau}b'\Omega e_{2}}{\Xi_{22}} \mathcal{Z}_{1}.
\end{equation}
The difference between this expression and the asymptotic distribution for the
minimum distance estimator subject to the positive definiteness condition is
that the term $\max(\mathcal{Z}_{1},0)$ in Equation~\ref{eq:ure:boundary-distro}
has been replaced by $\mathcal{Z}_{1}$. \citet[Section 4]{lp70} were the first
ones to point out that this increases the asymptotic mean squared error.

There are several possible approaches to inference on $\beta$ using
$\hat{\beta}$. I discuss two of them \citep[see][for a discussion of the
bootstrap and subsampling]{andrews99}. The first approach is based on the
observation that the conventional asymptotic standard errors based on the
assumption that no parameters are on the boundary (i.e.\ standard errors for
$\h{\beta}{umd}$) yield conservative confidence intervals when, in fact $\Xi$ is
reduced rank \citep[p.~1369]{andrews99}. The second approach suggested by
\citet{andrews99} is to do a pre-test of the hypothesis
$H_{0}\colon\Xi_{11}=\Xi_{22}\beta^{2}$ against
$H_{1}\colon\Xi_{11}>\Xi_{22}\beta^{2}$ to determine if the true parameter
$\Xi_{11}$ is at the boundary with critical values chosen such that the pre-test
is consistent as $n\to\infty$. If the test rejects, then we conclude that we're
not at the boundary, and we use \umd\ standard errors. Otherwise, we assume that
we're at the boundary, and, we use the asymptotic
distribution~\eqref{eq:ure:boundary-distro} to obtain confidence intervals.
Quantiles of the limiting distribution in \Cref{eq:ure:boundary-distro} can be
obtained by simulating draws of $\mathcal{Z}_{1}$ and $\mathcal{Z}_{2}$. The
pre-test used in this approach is, in fact, equivalent to a consistent test of
overidentifying restrictions, so that the modified Cragg-Donald test can be
used.

\subsection{Additional details for
  \texorpdfstring{\Cref{sec:specification-tests}}{Section 6}}
I first derive the expression for $\hat{J}_{\md}$. First, observe that, since
$\Qt(\hat{\beta}_{\re},S)=m_{\max}$,
\begin{multline}\label{eq:md:Wre:objective:max}
  \min_{\Xi_{11}=\Xi_{22}\beta^{2}}
  \mathcal{Q}_{n}(\beta,\Xi_{11},{\Xi}_{22},\hat{W}_{\re})
  =    \mathcal{Q}_{n}(\h{\beta}{re},\hat{\Xi}_{22,\re},\hat{W}_{\re})\\
  = \trace\left(((\Ni/n)I_{2}-S^{-1}T)^{2}\right)-(m_{\max}-\Ni/n)^{2}.
\end{multline}
Since
$\trace\left(((\Ni/n)I_{2}-S^{-1}T)^{2}\right)=(m_{\max}-\Ni/n)^{2}+(m_{\min}-\Ni/n)^{2}$,
it follows that~\eqref{eq:md:Wre:objective:max} can be written as
\begin{equation*}
  \mathcal{Q}_{n}(\h{\beta}{re},\hat{\Xi}_{22,\re},\hat{W}_{\re})=
  (m_{\min}-\Ni/n)^{2}.
\end{equation*}
It follows from the results in \Cref{sec:addit-deta-crefs} that if $S-(\Ni/n)T$
is not positive semi-definite (which is equivalent to $m_{\min}< k_{n}/n$), then
\begin{equation*}
  \min_{\Xi_{11}\geq\Xi_{22}\beta^{2}}\mathcal{Q}_{n}(\beta,\Xi_{11},\Xi_{22};\h{W}{re})=
  \min_{\Xi_{11}=\Xi_{22}\beta^{2}}\mathcal{Q}_{n}(\beta,\Xi_{11},{\Xi}_{22},\hat{W}_{\re}).
\end{equation*}
Otherwise,
$\min_{\Xi_{11}\geq\Xi_{22}\beta^{2}}\mathcal{Q}_{n}(\beta,\Xi_{11},\Xi_{22};\h{W}{re})=0$,
which yields the expression for $\hat{J}_{\md}$ as stated in the main text.

Next, I derive the asymptotic properties of overidentification tests proposed by
\citet{sargan58}, \citet{cd93}, and \citet{ar49}. Let
\begin{equation*}
  \h{J}{s}=\frac{\h{b}{liml}'T \h{b}{liml}}{\h{b}{liml}'(T-(\Ni/n)S) \h{b}{liml}}=
  \frac{m_{\min}}{1-\Ni/n-\Ne/n+ m_{\min}}
\end{equation*}
The \citet{sargan58} test rejects whenever
$n\h{J}{s}>q^{\chi^{2}_{k-1}}_{1-\ns}$, the $1-\ns$ quantile of a
$\chi^{2}_{k-1}$ distribution where $\ns$ denotes the desired nominal size. The
generalized likelihood ratio test based on the limited information likelihood of
\citet{ar49} replaces $\h{J}{s}$ with $\h{J}{ar}= \log(nm_{\min}/(\nus)+1)$, and
the \citet{cd93} test uses $\h{J}{cd}=m_{\min}$.


All three tests are equivalent in the sense that they all reject for large
values of $m_{\min}$. Therefore, the only difference between them in finite
samples is how well the chi-squared approximation controls size in each case.
While under standard asymptotics their asymptotic distributions coincide and
therefore do not provide any guidance as to which test has the best size
control, allowing for $\alphai,\alphae>0$ reverses this conclusion.
\begin{lemma}\label{th:jk:distro}
  Under \Cref{an:pr,,an:mi,an:rc}, if $\Ni\to\infty$,
  \begin{align*}
    \textstyle \Prob\left(n\h{J}{s}\geq
    q^{\chi^{2}_{\Ni-1}}_{1-\ns}\right)
    & \to
      \begin{cases}
\Phi\left(
        \frac{ \Phi^{-1}(\ns)}{
          \sqrt{(1-\alphai)(1+(1-\alphai)\kappa\delta/2 )}}\right)
        & \text{if $\alphae=0$,} \\
        1
        & \text{otherwise.}\\
      \end{cases}\\
    \Prob\left(n\h{J}{cd}\geq q^{\chi^{2}_{\Ni-1}}_{1-\ns}\right)
    &\to\Phi\left(\frac{\Phi^{-1}(\ns)}{\sqrt{(1-\alphae)/(\alphanu)
      +\kappa\delta/2}} \right),
  \end{align*}
  and if $\alphai>0$, then
  $\Prob\left(n\h{J}{ar}\geq q^{\chi^{2}_{\Ni-1}}_{1-\ns}\right) \to 1$, where
  $\Phi(\cdot)$ is the cdf of a standard normal distribution.
\end{lemma}
\begin{sproof}[\Cref{th:jk:distro}]
  Let $\tilde{\alpha}=(1-\alphae)/(1-\alphai-\alphae)$. By \Cref{th:overid} and
  the delta method,
  \begin{align*}
    \textstyle \frac{n}{\sqrt{\Ni}}
    \left(\h{J}{s}-\frac{\alphai}{1-\alphae} \right)
    & \textstyle\indist \mathcal{N}\left(0,\frac{
      2\tilde{\alpha}+\kappa\delta }{(1-\alphae
      )^{2}\tilde{\alpha}^{2}}
      \right)\\
    \textstyle
    \frac{n}{\sqrt{\Ni}}\left(\h{J}{ar}-\log(\tilde{\alpha})\right)
    &\textstyle\indist
      \mathcal{N}\left(0,\frac{2\tilde{\alpha}+
      \kappa\delta}{(1-\alphae)^{2}}\right)\\
    \textstyle
    \frac{n}{\sqrt{\Ni}}
    \left(\h{J}{cd}-\alphai\right)
    & \textstyle\indist\mathcal{N}(0,2\tilde{\alpha}+\delta\kappa),
  \end{align*}
  I use the approximation from \citet{peiser43} that as $k\to \infty$,
  \begin{equation*}
    q_{1-\ns}^{\chi^{2}_{k}}=k+\Phi^{-1}(1-\ns)\sqrt{2k}+O(1).
  \end{equation*}
  Therefore if $\tau>0$,
  \begin{multline*}
    \Prob\left(n\h{J}{cd}\geq q^{\chi^{2}_{\Ni-1}}_{1-\ns}\right)=
    \Prob\left(\frac{n}{\sqrt{\Ni}}(\h{J}{cd}-\alphai)\geq
      \Phi^{-1}(1-\ns)\sqrt{2}+O(1/\sqrt{\Ni})\right)\\
    = \Prob\left(\mathcal{N}(0,1)+o_{p}(1)\geq
      \frac{\Phi^{-1}(1-\ns)}{\sqrt{\tilde{\alpha}+\kappa\delta/2}}+o(1)\right)
    \to\Phi\left(\frac{\Phi^{-1}(\ns)}{\sqrt{\tilde{\alpha}+\kappa\delta/2}}
    \right).
  \end{multline*}
  Similarly,
  \begin{equation*}
    \begin{split}
      \Prob\left(n\h{J}{s}\geq q^{\chi^{2}_{\Ni-1}}_{\ns}\right)&=
      \Prob\left(n\h{J}{s}\geq
        \Ni+\Phi^{-1}(1-\ns)\sqrt{2\Ni}+O(1)\right)\\
      &= \Prob\left( \sqrt{\frac{
      2\tilde{\alpha}+\kappa\delta }{(1-\alphae
      )^{2}\tilde{\alpha}^{2}}}\mathcal{N}(0,1)+o_{p}(1) \geq
        -\frac{\sqrt{\Ni}\alphae}{(1-\alphae)}
        +\Phi^{-1}(1-\ns)\sqrt{2}+o(1)\right).
    \end{split}
  \end{equation*}
  Now, if $\alphae>0$, then the right-hand side converges to $-\infty$, so that
  the rejection probability converges to one. If $\alphae=0$, then
  \begin{equation*}
      \Prob\left(n\h{J}{s}\geq q^{\chi^{2}_{\Ni-1}}_{\ns}\right)\to\Phi\left(
        \frac{ \Phi^{-1}(\ns)}{
          \sqrt{(1-\alphai)(1+(1-\alphai)\kappa\delta/2 )}}\right).
  \end{equation*}
  Finally,
  \begin{equation*}
    \begin{split}
      \Prob\left(n\h{J}{ar}\geq q^{\chi^{2}_{\Ni-1}}_{\ns}\right)&=
      \Prob\left(\frac{n}{\sqrt{\Ni}}(\h{J}{ar}-n\log(\tilde{\alpha}))\geq
        -\frac{n}{\sqrt{\Ni}}\log(\tilde{\alpha})+
        \sqrt{\Ni}+\Phi^{-1}(1-\ns)\sqrt{2}+o(1)\right)\\
      &= \Prob\left( \frac{\sqrt{2\tilde{\alpha}+
            \kappa\delta}}{1-\alphae}\mathcal{N}(0,1)+o_{p}(1)\geq
        \frac{n}{\sqrt{\Ni}}\left(\Ni/n-\log(\tilde{\alpha})\right)
        +\Phi^{-1}(1-\ns)\sqrt{2}+o(1)\right).
    \end{split}
  \end{equation*}
  Since $\alphai\leq-\log(1-\alphai)$,
  \begin{equation*}
    \alphai-\log(\tilde{\alpha})
    \leq
    \log\left(\frac{1}{1-\alphai}\right)
    -\log\left(\frac{1-\alphae}{\alphanu}\right)
    =\log\left(\frac{1-\alphai-\alphae}{(1-\alphai)(1-\alphae)}\right)
    \leq 0,
  \end{equation*}
  with equality only if $\alphai=0$, so that the right-hand side of the previous
  display converges to $-\infty$ if $\alphai>0$.
\end{sproof}

\section{Proof of \texorpdfstring{\Cref{th:clt}}{Lemma A.2}}
\begin{sproof}
  Let $K_{d}=2N_{d}-I_{d^{2}}$ denote the commutation matrix, which has the
  property that $K_{d}\mkvec(A)=\mkvec(A')$, where $A$ is a $d\times d$ matrix.
  To show part~\ref{it:variance}, note that for any
  $v_{1},v_{2},v_{3},v_{4}\in\mathbb{R}^{d}$,
  \begin{equation}
    v_{1}v_{2}'\otimes v_{3}v_{4}'=K_{d}(v_{3}v_{2}'\otimes v_{1}v_{4}').\label{eq:vec-Kron2}
  \end{equation}
  This follows from
  $v_{1}v_{2}'\otimes v_{3}v_{4}'=v_{1}\otimes (v_{3}v_{2}'\otimes
  v_{4}')=K_{d}(v_{3}v_{2}'\otimes v_{4}')\otimes v_{1}$, where the second
  equality uses the identity $K_{d}(A\otimes v)=v\otimes A$ for any
  $A\in\mathbb{R}^{d\times d'}$ and $v\in\mathbb{R}^{d}$ \citep[Theorem
  3.1(ix)]{mn79}. Furthermore,
  \begin{subequations}
    \begin{align}
      \mkvec(\Qf)
      &=(I_{G^{2}}+K_{G})(I_{G}\otimes \Mm'\Pm)\mkvec(\Em)+\mkvec(\Em'\Pm\Em)+
        \mkvec(\Mm'\Pm\Mm),\label{eq:q-a}\\
      \E[\Em'\Pm\Em]
      &=\trace(\Pm)\Omega_{n},\label{eq:q-b}\\
      \E[\mkvec(\Qf)]
      &=\mkvec(\Mm'\Pm\Mm+\trace(\Pm)\Omega_{n}),\label{eq:q-c}\\
      \E[\mkvec(\Em)\mkvec(\Em)']
      &=\Omega_{n}\otimes I_{n},\label{eq:q-d}\\
      \E[\mkvec(\Em)\mkvec(\Em'\Pm\Em)']
      &=\E[\ei{i}\otimes \diag(\Pm)
        \otimes \ei{i}'\otimes\ei{i}'],\label{eq:q-e}\\
      \E\mkvec(\Em'\Pm\Em)\mkvec(\Em'\Pm\Em)'
      &=\delta_{n}\E[\ei{i}\ei{i}'\otimes \ei{i}\ei{i}']
        +(\trace(\Pm)^{2}-\delta_{n})\mkvec(\Omega_{n})\mkvec(\Omega_{n})'
        \label{eq:q-ff}\\
      &\qquad+(\trace(\Pm^{2})-\delta_{n})(I_{G^{2}}+K_{G})\Omega_{n}\otimes
        \Omega_{n}
        \nonumber
    \end{align}
  \end{subequations}
  where~\eqref{eq:q-a} follows by the definition of the commutation matrix,
  \eqref{eq:q-b} follows from the expansion
  $\Em'\Pm\Em=\sum_{i,j}p_{ij}\ei{i}\ei{j}'$, \eqref{eq:q-ff} also follows from
  this expansion and from~\eqref{eq:vec-Kron2}, \eqref{eq:q-c} follows
  from~\eqref{eq:q-b}. \Cref{eq:q-d,eq:q-e} follow by direct calculation.
  Therefore,
  \begin{align*}
    \var[\mkvec(\Qf)]
    &=
      (I_{G^{2}}+K_{G})(I_{G}\otimes \Mm'\Pm)\E[\mkvec(\Em)
      \mkvec(\Em)'](I_{G}\otimes \Pm\Mm)(I_{G^{2}}+K_{G})\\
    &\qquad+      (I_{G^{2}}+K_{G})(I_{G}\otimes \Mm'\Pm)\E[\mkvec(\Em)
      \mkvec(\Em'\Pm\Em)']\\
    &\qquad+\E[\mkvec(\Em'\Pm\Em)\mkvec(\Em)']
      (I_{G}\otimes \Pm\Mm)(I_{G^{2}}+K_{G})
      +\E\mkvec(\Em'\Pm\Em)\mkvec(\Em'\Pm\Em)'\\
    &\qquad-\trace(\Pm)^{2}\mkvec(\Omega_{n})\mkvec(\Omega_{n})'\\
    &=(I_{G^{2}}+K_{G})(\Omega_{n}\otimes \Psi_{n})(I_{G^{2}}+K_{G})
      +\trace(\Pm^{2})(I_{G^{2}}+K_{GG})\Omega_{n}\otimes \Omega_{n}\\
    &\qquad+(I_{G^{2}}+K_{G})\E[\ei{i}\ei{i}'\otimes \overbar{m}_{n}\ei{i}']
      +\E[\ei{i}\ei{i}'\otimes \ei{i}\overbar{m}_{n}'](I_{G^{2}}+K_{G})\\
    &\qquad      +\delta_{n}\left(\E[\ei{i}\ei{i}'\otimes \ei{i}\ei{i}']
      -\mkvec(\Omega_{n})\mkvec(\Omega_{n})'
      -(I_{G^{2}}+K_{GG})\Omega_{n}\otimes \Omega_{n}
      \right),
  \end{align*}
  where the first equality uses~\eqref{eq:q-a}--\eqref{eq:q-c}, and the second
  equality uses~\eqref{eq:q-d}--\eqref{eq:q-ff}. The result then follows by
  applying the identities~\eqref{eq:vec-Kron2} and
  $A_{1}\otimes A_{2}=K_{d}(A_{2}\otimes A_{1})$ for any
  $A_{1},A_{2}\in\mathbb{R}^{d\times d}$ \citep[Theorem 3.1(ix)]{mn79}.

  The proof of part~\ref{it:centr-limit-theor} adapts the arguments in
  \citet{cshnw12} and \citet{hhn08}. By the Cramér–Wold device, it suffices to
  prove the result for
  \begin{equation*}
    \mkvec(A)'\mkvec(\Qf)=\trace(A'\Qf)
  \end{equation*}
  where $A\in\mathbb{R}^{G\times G}$ is an arbitrary matrix of constants. Since
  $\Qf$ is symmetric, we can without loss of generality assume that $A$ is also
  symmetric. Expanding the expression, and using symmetry of $\Pm$ yields
  \begin{equation*}
    \begin{split}
      \trace(A\Qf-\E[A\Qf]) &= \sum_{i=1}^{n}\sum_{j=1}^{n}(\mi{j}+\ei{j})'
      A(\mi{i}+\ei{i}) p_{ij}-\sum_{i=1}^{n}\sum_{j=1}^{n}\mi{j}' A
      \mi{i}p_{ij}-
      \sum_{i=1}^{n}p_{ii}\trace(A\Omega_{n})\\
      &= \sum_{i=1}^{n}W_{in} +\sum_{i=2}^{n}\sum_{j=1}^{i-1} 2 p_{ij} \ei{i}'A
      \ei{j}=\sum_{i=1}^{n}y_{in},
    \end{split}
  \end{equation*}
  where $y_{in}=W_{in}+2S_{in}$ for $i\geq 2$, $y_{1n}=W_{1n}$ and
  \begin{align*}
    W_{in}&=2e_{in}'\Pm\Mm A \ei{i}+p_{ii}(\ei{i}'A \ei{i}-\trace(A\Omega_{n})),\\
    S_{in}&=\sum_{j=1}^{i-1}p_{ij}\ei{i}'A\ei{j},
  \end{align*}

  Note that $y_{in}$ is a martingale difference array with respect to the
  filtration $\mathcal{F}_{in}=\sigma(\ei{1},\dotsc,\ei{i-1,})$. By the
  martingale central limit theorem, it therefore suffices to show that for some
  $\varepsilon>0$,
  \begin{equation}
    \label{eq:lyapunov}  \sum_{i=1}^{n}\E[\abs{y_{in}}^{2+\varepsilon}]=o(1),
  \end{equation}
  and that the conditional variance
  $\sum_{i=1}^{n}\CE{y_{in}^{2}}{\mathcal{F}_{i-1,n}}$ converges. By the Loève
  $c_{r}$-inequality if
  \begin{align}
    \label{eq:linde-1} \E[\abs{\ei{i}'A \ei{i}-\trace(A\Omega_{n})}^{4}]
    \sum_{i=1}^{n}p_{ii}^{4}&=o(1),\\
    \label{eq:linde-2}
    \sum_{i=2}^{n}\E[S_{in}^{4}]&=o(1),\qquad\text{and}\\
    \label{eq:linde-3}
    \sum_{i=1}^{n}\E[(e_{in}'\Pm M_{n}A\ei{i})^{4}]&=o(1),
  \end{align}
  then~\eqref{eq:lyapunov} holds with $\varepsilon=2$. Now, \eqref{eq:linde-1}
  follows from Assumptions~\ref{it:four-plus-delta-moments}
  and~\ref{it:P-conditions}. To show~\eqref{eq:linde-2}, note that expanding the
  expression yields
  \begin{equation*}
    \sum_{i=2}^{n}\E[S_{in}^{4}]
    =2\sum_{i=2}^{n}\sum_{j=1}^{i-1}\sum_{k=1}^{i-1}
    p_{ij}^{2}p_{ik}^{2} \E[ (\ei{j}'A \ei{i})^{2}(\ei{i}'A \ei{k})^{2}]\leq C\sum_{i=2}^{n}\sum_{j=1}^{i-1}\sum_{k=1}^{i-1} p_{ij}^{2}p_{ik}^{2}
    \leq C\sum_{i=1}^{n}\left(\sum_{j=1}^{n} p_{ij}^{2} \right)^{2},
  \end{equation*}
  for some constant $C$, which is $o(1)$ by Assumption~\ref{it:P-conditions}.
  Next, to show~\eqref{eq:linde-3}, note that
  \begin{equation*}
    \sum_{i=1}^{n}\E[(e_{in}'\Pm \Mm Au_{in})^{4}]\leq
    \E[ \norm{Au_{in}}^{4}]\sum_{i=1}^{n}\norm{e_{in}'\Pm \Mm}^{4},
  \end{equation*}
  which is also $o(1)$ by Assumptions~\ref{it:four-plus-delta-moments}
  and~\ref{it:M-condition}.

  It remains to show convergence of the conditional variance. By
  Assumption~\ref{it:convergences}, it suffices to show that
  \begin{equation}
    \label{eq:cond-variance}
    \sum_{i=1}^{n}\CE{y_{in}^{2}}{\mathcal{F}_{i-1,n}}-\var(\trace(A\Qf))\inprob 0.
  \end{equation}
  Since $\CE{W_{in}^{2}}{\mathcal{F}_{i-1,n}}=\E[W_{in}^{2}]$, and since
  $ \var(\trace(A\Qf))=\sum_{i=1}^{n}\E[W_{in}^{2}]+
  4\sum_{i=2}^{n}\E[S_{in}^{2}]$, the left-hand side of~\eqref{eq:cond-variance}
  can be written as
  \begin{equation}\label{eq:cond-variance-simplified}
    \sum_{i=1}^{n}\CE{y_{in}^{2}}{\mathcal{F}_{i-1,n}}-\var(\trace(A\Qf))=
    4\sum_{i=2}^{n}\left(
      \CE{S_{in}^{2}}{\mathcal{F}_{i-1,n}}-
      \E[S_{in}^{2}]
    \right)+
    4\sum_{i=2}^{n}\CE{W_{in}S_{in}}{\mathcal{F}_{i-1,n}}.
  \end{equation}
  Letting $\Pm^{L}$ denote the lower triangular matrix with elements
  $p_{ij}1\{i>j\}$, we can write the second sum
  in~\eqref{eq:cond-variance-simplified} as
  \begin{equation*}
    \sum_{i=2}^{n}\CE{W_{in}S_{in}}{\mathcal{F}_{i-1}}
    =\sum_{i=1}^{n}\sum_{j=1}^{n}(\Pm^{L})_{ij}
    \E[W_{in}\ei{i}'] A\ei{j}
    =\trace\left(U'{\Pm^{L}}'\E[D_{W}U]A\right)
    =\sum_{g=1}^{G}e_{gG}'U'{\Pm^{L}}' \overbar{w}_{g},
  \end{equation*}
  where $D_{W}$ denotes a diagonal matrix with elements $(D_{W})_{ii}=W_{in}$,
  and $\overbar{w}_{g}=\E[D_{W}U]A e_{gG}$. The variance of the summand on the
  right-hand side is given by
  \begin{equation*}
    \begin{split}
      \E[(e_{g}'U'{\Pm^{L}}'\overbar{w}_{g})^{2}]&=
      \Omega_{gg}\overbar{w}_{g}'{\Pm^{L}}{\Pm^{L}}'\overbar{w}_{g}
      \leq\Omega_{gg}\norm{{\Pm^{L}}{\Pm^{L}}'}_{F}\norm{\overbar{w}_{g}}^{2}\\
      & =\Omega_{gg}\norm{{\Pm^{L}}{\Pm^{L}}'}_{F}
      \sum_{i=1}^{n}\E[W_{in}\ei{i}'Ae_{g}]^{2} \leq \Omega_{gg}
      e_{g}'A\Omega_{n} Ae_{g} \norm{{\Pm^{L}}{\Pm^{L}}'}_{F}
      \sum_{i=1}^{n}\E[W_{in}^{2}],
    \end{split}
  \end{equation*}
  where $\norm{\cdot}_{F}$ denotes the Frobenius norm. It follows by Loève
  $C_{r}$ inequality and Assumption~\ref{it:convergences} that
  $\sum_{i=1}^{n}\E[W_{in}^{2}]$ is bounded, and
  \begin{multline*}
    \norm{{\Pm^{L}}{\Pm^{L}}'}_{F}^{2}
    = \sum_{k<i}p_{ik}^{4}+ 2\sum_{k<\ell<i}p_{ik}^{2}p_{i\ell}^{2}+
    2\sum_{k<i<j}p_{ik}^{2}p_{jk}^{2}+
    4\sum_{k<\ell<i<j}p_{ik}p_{jk}p_{i\ell}p_{j\ell}\\
    \leq 5\sum_{i}(\sum_{j}p_{ij}^{2})^{2}+
    4\sum_{k<\ell<i<j}p_{ik}p_{jk}p_{i\ell}p_{j\ell} =o(1),
  \end{multline*}
  where the last equality follows by Assumption~\ref{it:P-conditions}. Hence, by
  Markov inequality, the second term in~\eqref{eq:cond-variance-simplified} is
  $o_{p}(1)$. Next, consider the first term
  in~\eqref{eq:cond-variance-simplified}, which can be written as
  \begin{equation}\label{eq:first-term-condvar}
    \begin{split}
      \sum_{i=2}^{n}\left( \CE{S_{in}^{2}}{\mathcal{F}_{i-1,n}}- \E[S_{in}^{2}]
      \right)& = \sum_{i=2}^{n}
      \left(\sum_{k=1}^{i-1}\sum_{j=1}^{i-1}p_{ij}p_{ik}\ei{k}'A\Omega A \ei{j}
        -\sum_{j=1}^{i-1}p_{ij}^{2} \trace(\Omega A\Omega A) \right)\\
      &= \sum_{i=2}^{n}\sum_{j=1}^{i-1}p_{ij}^{2}(\ei{j}'A\Omega A
      \ei{j}-\trace(\Omega A\Omega A))
      +2\sum_{i=2}^{n}\sum_{k=1}^{i-1}\sum_{j=1}^{k-1}p_{ij}p_{ik}\ei{k}'A\Omega
      A \ei{j}.
    \end{split}
  \end{equation}
  Variance of the first term in~\eqref{eq:first-term-condvar} is given by
  \begin{equation*}
    \begin{split}
      \var\left( \sum_{i=2}^{n}\sum_{j=1}^{i-1}p_{ij}^{2}(\ei{j}'A\Omega A
        \ei{j}-\trace(\Omega A\Omega A))\right)&= \E[(\ei{i}'A\Omega A
      \ei{i}-\trace(\Omega A\Omega A))^{2}] \sum_{i=2}^{n}\sum_{j=1}^{i-1}
      \sum_{\ell=j+1}^{n} p_{\ell j}^{2} p_{ij}^{2},
    \end{split}
  \end{equation*}
  which converges to zero since the triple sum
  $\sum_{i=2}^{n}\sum_{j=1}^{i-1} \sum_{\ell=j+1}^{n} p_{\ell j}^{2} p_{ij}^{2}$
  is bounded by
  \begin{equation}\label{eq:triple-bound}
    \sum_{i=1}^{n}\sum_{j=1}^{n}
    \sum_{\ell=1}^{n} p_{\ell j}^{2}
    p_{ij}^{2}=\sum_{i=1}^{n}\left(\sum_{j=1}^{n}p_{ij}^{2}\right)^{2}=o(1),
  \end{equation}
  where the last equality follows by Assumption~\ref{it:P-conditions}. Variance
  of the second term in~\eqref{eq:first-term-condvar} given by
  \begin{equation*}
    \begin{split}
      \var\left(
        \sum_{i=2}^{n}\sum_{k=1}^{i-1}\sum_{j=1}^{k-1}p_{ij}p_{ik}\ei{k}'A\Omega
        A \ei{j} \right) & = \trace((A\Omega)^{4})
      \sum_{i=2}^{n}\sum_{k=1}^{i-1}\sum_{j=1}^{k-1}\sum_{\ell=k+1}^{n}
      p_{ij}p_{ik} p_{\ell j}p_{\ell k}\\
      & = \trace((A\Omega)^{4})\left[ \sum_{j<k<i} p_{ij}^{2}p_{ik}^{2} +
        2\sum_{j<k<i<\ell}^{} p_{ji}p_{ik} p_{\ell j}p_{\ell k} \right],
    \end{split}
  \end{equation*}
  where the first sum is again bounded by~\eqref{eq:triple-bound}, and the
  second term equals
  $\sum_{i<j<k<\ell}^{} p_{ik}p_{i \ell } p_{jk} p_{j \ell }$, which is $o(1)$
  by Assumption~\ref{it:P-conditions}. Therefore, by Markov inequality, the
  first term in~\eqref{eq:cond-variance-simplified} is $o_{p}(1)$, so
  that~\eqref{eq:cond-variance} holds, which proves the theorem.
\end{sproof}

\bibliographystyle{ecca}%
\renewcommand{\bibfont}{\footnotesize}%
\bibliography{many-iv-library}